\documentclass[aps,floats]{revtex4-2}
\usepackage{amsmath,amssymb}
\usepackage{graphicx,epsfig}
\usepackage[greek,english]{babel}
\usepackage{bbold}

\begin{document}
\bibliographystyle {plain}

\pdfoutput=1
\def\oppropto{\mathop{\propto}} 
\def\opsimeq{\mathop{\simeq}}
\def\opoverderline{\mathop{\overline}}
\def\operarrow{\mathop{\longrightarrow}}
\def\opsim{\mathop{\sim}}

\def\opmin{\mathop{\min}} 
\def\opmax{\mathop{\max}} 
\def\oplim{\mathop{\lim}}

%%%%%%%%%%%%%%%%%%%%%%%%%%%%%%%%%%%%%%%%%%%%%%%%%%%%%%%%%%%%%%%%%%%%%%%%%%%%
\title{  Markov generators as non-hermitian supersymmetric quantum Hamiltonians: \\
spectral properties via bi-orthogonal basis and Singular Value Decompositions } 

%%%%%%%%%%%%%%%%%%%%%%%%%%%%%%%%%%%%%%%%%%%%%%%%%%%%%%%%%%%%%%%%%%%%%%%%%%%%

\author{C\'ecile Monthus}
\affiliation{Universit\'e Paris-Saclay, CNRS, CEA, Institut de Physique Th\'eorique, 91191 Gif-sur-Yvette, France}

%%%%%%%%%%%%%%%%%%%%%%%%%%%%%%%%%%%%%%%%%%%%%%%%%%%%%%%%%%%%%%%%%%%%%%%%%%%%

\begin{abstract}
Continuity equations associated to continuous-time Markov processes can be considered as Euclidean Schr\"odinger equations, where the non-hermitian quantum Hamiltonian $\bold{H}={\bold{div}}{\bold J}$ is naturally factorized into the product of the divergence operator ${\bold {div}}$ and the current operator ${\bold J}$. For non-equilibrium Markov jump processes in a space of $N$ configurations with $M$ links and $C=M-(N-1)\geq 1$ independent cycles, this factorization of the $N \times N$ Hamiltonian ${\bold H}={\bold I}^{\dagger}{\bold J}$ involves the incidence matrix ${\bold I}$ and the current matrix ${\bold J}$ of size $M \times N$, so that the supersymmetric partner ${\hat{\bold H}}= {\bold J}{\bold I}^{\dagger}$ governing the dynamics of the currents living on the $M$ links is of size $M \times M$. To better understand the relations between the spectral decompositions of these two Hamiltonians $\bold{H}={\bold I}^{\dagger}{\bold J}$ and ${\hat {\bold H}} ={\bold J}{\bold I}^{\dagger}$ with respect to their bi-orthogonal basis of right and left eigenvectors that characterize the relaxation dynamics towards the steady state and the steady currents, it is useful to analyze the properties of the Singular Value Decompositions of the two rectangular matrices ${\bold I}$ and ${\bold J} $ of size $M \times N$ and the interpretations in terms of discrete Helmholtz decompositions. This general framework concerning Markov jump processes can be adapted to non-equilibrium diffusion processes governed by Fokker-Planck equations in dimension $d$, where the number $N$ of configurations, the number $M$ of links and the number $C=M-(N-1)$ of independent cycles become infinite, while the two matrices ${\bold I}$ and ${\bold J}$ become first-order differential operators acting on scalar functions to produce vector fields.

\end{abstract}

\maketitle

%%%%%%%%%%%%%%%%%%%%%%%%%%%%%%%%%%%%%%%%%%%

\section{ Introduction }

While Markov processes satisfying detailed-balance are well understood,
in particular via the similarity transformation of the opposite Markov generator into a supersymmetric quantum Hamiltonian of the form $H=Q^{\dagger} Q$ with a real energy spectrum 
(see the textbooks \cite{gardiner,vankampen,risken}  and the applications to various models \cite{glauber,Felderhof,siggia,kimball,peschel,jpb_antoine,pierre,texier,us_eigenvaluemethod,Castelnovo,c_pearson,c_boundarydriven,c_largeDevsusy}), non-equilibrium Markov processes involving steady currents 
remain challenging and require the introduction of many new ideas 
(see the reviews with different scopes \cite{derrida-lecture,harris_Schu,searles,harris,mft,sollich_review,lazarescu_companion,lazarescu_generic,jack_review}, 
the PhD Theses \cite{fortelle_thesis,vivien_thesis,chetrite_thesis,wynants_thesis,chabane_thesis,duBuisson_thesis},
 the Habilitation Thesis \cite{chetrite_HDR}, and the recent lecture notes \cite{luxembourg}).
 
 Among the various perspectives that have emerged to analyze non-equilibrium Markov jump processes, 
 the appropriate graph theory based on the $N$ configurations, on the $M$ links existing between them, 
 and on the $C=M-(N-1)$ independent cycles
 has proven to be very useful 
 (see \cite{luxembourg,phdPolettini,CycleCocycle,gottingen,maes2022,lecomte} and references therein)
  since the pioneering work of Schnakenberg \cite{Schnakenberg}.
  In the present paper, we will keep the idea that the rectangular incidence matrix ${\bold I}$ of size $M \times N$ between links and sites plays an essential role, with its adjoint ${\bold I}^{\dagger} = {\bold {div}} $ that represents the discrete divergence.
  But we will consider on the same footing the rectangular current matrix ${\bold J}$ of size $M \times N$,
  that can be considered as a deformation of the incidence matrix ${\bold I}$ via the transition rates.
 These two matrices appear naturally in the rewriting of the opposite Markov generator as the non-hermitian supersymmetric Hamiltonian $ \bold{H}= {\bold I}^{\dagger}  {\bold J} $ of size $N \times N$ that governs the dynamics of the probability density living on the $N$ sites, while the 
 supersymmetric partner ${\hat {\bold H} } = {\bold J}  {\bold I}^{\dagger}$ of size $M \times M$ governs the dynamics of the currents living on the $M$ links.
 
 The main goal of the present paper is to analyze the consequences of these supersymmetric factorizations for the spectral properties of the non-hermitian Hamiltonians $ \bold{H}= {\bold I}^{\dagger}  {\bold J} $ and ${\hat {\bold H} } = {\bold J}  {\bold I}^{\dagger}$ : we will first focus on their spectral decompositions 
  with respect to their bi-orthogonal basis of right and left eigenvectors that characterize the relaxation dynamics towards the steady state and the steady currents, and we will then describe
 how many important properties can be understood via the Singular Value Decompositions of the two rectangular matrices ${\bold I}$ and ${\bold J} $
  and the interpretations in terms of discrete Helmholtz decompositions.
  Note that independently of the present motivations coming from Markov processes,
  the field of non-hermitian physics has become relevant in many areas 
 (see the review \cite{ReviewNonHermitian} and references therein),
 while the interest into hermitian supersymmetric quantum hamiltonians of the form $H=Q^{\dagger} Q$
 (see the review \cite{review_susyquantum} and references therein)
 has expanded towards various non-hermitian cases  \cite{ioffe,rosas,bagarello,cruz}.
 
The paper is organized as follows :

$\bullet$ Most sections of the main text are devoted to the general analysis of non-equilibrium Markov jump processes in a space of $N$ configurations :

 In section \ref{sec_jump}, we recall that the Markov generator can be considered as the opposite of a non-hermitian quantum Hamiltonian ${\bold H}$ of size $N \times N$, 
and that the spectral decomposition in the bi-orthogonal basis
of its right and left eigenvectors is useful to analyze the convergence of 
configurations probabilities and configurations observables towards their steady values.

 In section \ref{sec_current}, we describe the properties of the currents 
that are defined on the $M$ links
existing between the $N$ configurations : we analyze the consequences of the supersymmetric factorization of the $N \times N$ non-hermitian Hamiltonian $ \bold{H}= {\bold I}^{\dagger}  {\bold J} $ in terms of the incidence matrix ${\bold I}$
 and the current matrix ${\bold J} $ that are both of size $M \times N$, 
 so that the supersymmetric partner 
${\hat {\bold H} } = {\bold J}  {\bold I}^{\dagger}$ governing the dynamics of the currents is a priori of size $M \times M$, although the physical currents live on a smaller subspace of size $N$.

 In order to clarify this last point, we discuss the properties of the Singular Value Decompositions of the two rectangular matrices ${\bold I}$ and ${\bold J} $ in sections \ref{sec_SVDI} and \ref{sec_SVDJ} respectively. 

This general perspective is illustrated by a simple translation-invariant example in Appendix \ref{app_example}.

$\bullet$ In the last section section \ref{app_diffusion} of the main text, 
we describe how the previous framework concerning Markov jump processes 
can be adapted to non-equilibrium diffusion processes governed by Fokker-Planck generators in dimension $d=3$, even if there are some important technical differences : the number $N$ of configurations, the number $M$ of links and the number $C=M-(N-1) $ of independent cycles become infinite, while the two matrices ${\bold I}$ and ${\bold J} $ of size $M \times N $ become first-order differential operators acting on scalar functions to produce vector fields. 

Our conclusions are summarized in section \ref{sec_conclusion}.

%%%%%%%%%%%%%%%%%%%%%%%%%%%

\section{ Non-equilibrium Markov jump processes in a space of $N$ configurations }

\label{sec_jump}

In this section, we introduce the general notations for Markov jump processes in a space of $N$ configurations.
We recall that the Markov generator can be considered as the opposite of a non-hermitian quantum Hamiltonian ${\bold H}$ of size $N \times N$, and that its spectral decomposition in the bi-orthogonal basis
of right and left eigenvectors is useful to analyze the convergence of 
configurations probabilities and configurations observables towards their steady values.

\subsection{ Markov generator $w(.,.)$ of size $N \times N$ in the space of the $N$ configurations   }

 For a Markov jump process over $N$ configurations $x$, 
 the master equation for the probability $p_t(x) $ to be at configuration $x$ at time $t$
\begin{eqnarray}
 \partial_t p_t(x) && = \sum_{x' } w(x,x') p_t( x' ) = w(x,x)  p_t( x ) + \sum_{x' \ne x} w(x,x') p_t( x' )
\label{mastereq}
\end{eqnarray}
involves the $N \times N$ Markov matrix $w(.,.) $, 
where the off-diagonal element $w(x,x') \geq 0 $ represents the transition rate
from $x'$ towards $x$, while the diagonal element $w(x,x) $ is negative and determined by the off-diagonal elements
\begin{eqnarray}
w(x,x)  = - \sum_{x' \ne x} w(x',x) <0
\label{wdiag}
\end{eqnarray}
The steady state $p_*(x)$ satisfies
\begin{eqnarray}
0= \partial_t p_*(x) && = \sum_{x' } w(x,x') p_*( x' ) 
\label{mastereqsteady}
\end{eqnarray}

%%%%%%%%%%%%%%%%%%%%%%%%%%%%%%%%%%%%%%%%%%%

\subsection{ Parametrization of the transition rates $w(x,x')$    }

For each pair of configurations $(x \ne x')$ related by strictly positive transition rates $ w(x',x)  w(x,x') >0$ ,
 it is useful to introduce the parametrization
 \begin{eqnarray}
  w(x',x) && = D(x',x) e^{A(x',x)}
 \nonumber \\ 
 w(x,x') && = D(x',x) e^{- A(x',x)}
\label{wrates}
\end{eqnarray}
in terms of the positive symmetric function
 \begin{eqnarray}
D(x',x) = \sqrt{ w(x,x')  w(x',x) } = D(x,x') >0
\label{Dsym}
\end{eqnarray}
and of the antisymmetric function
 \begin{eqnarray}
A(x',x) = \frac{1}{2} \ln \left( \frac{ w(x',x) }{  w(x,x') } \right) = - A(x,x')
\label{Avecpot}
\end{eqnarray}

%%%%%%%%%%%%%%%%%%%%%%%%%%%%%%%%%%%%%%%%%%%%%%%%

\subsection{ Opposite of the Markov generator $(-{\bold w})={\bold H}$ as a non-hermitian quantum Hamiltonian ${\bold H} \ne {\bold H}^{\dagger}$   }

It is useful to consider that the master Eq. \ref{mastereq} is an Euclidean Schr\"odinger equation
\begin{eqnarray}
- \partial_t \vert p_t \rangle  = {\bold H} \vert p_t \rangle
\label{mastereqschro}
\end{eqnarray}
where the quantum Hamiltonian ${\bold H} $
 is simply the opposite of the Markov matrix ${\bold H} = - {\bold w}$.
 With the parametrization of Eq. \ref{wrates} for the transition rates,
 the off-diagonal elements read
 \begin{eqnarray}
 {\bold H} (x',x) && =- w(x',x) = - D(x',x) e^{ A(x',x)} 
   \nonumber \\ 
 {\bold H} (x,x') && =-  w(x,x') = - D(x',x) e^{-A(x',x)}
\label{Hwrates}
\end{eqnarray} 
The symmetric function $D(x',x)=D(x,x')$ of Eq. \ref{Dsym} represents 
the symmetric part of the hopping amplitude between the two sites $x$ and $x'$,
while the antisymmetric function $A(x,x') =- A(x',x)$ of Eq. \ref{Avecpot} plays the role of an imaginary vector potential
that is responsible for the non-hermitian character ${\bold H} \ne {\bold H}^{\dagger}$.
 This correspondence between continuous-time Markov generators and non-hermitian quantum Hamiltonians 
 involving imaginary vector potentials has been already emphasized for diffusion processes in arbitrary dimension \cite{us_gyrator} (see also the section \ref{app_diffusion} below for 
 the special case of dimension $d=3$)
 and for Markov jump processes on hypercubic lattices \cite{c_missing}.
Note that non-hermitian quantum Hamiltonians with imaginary vector potentials
have been much studied in various contexts since the pioneering works of Hatano and Nelson \cite{Hatano,nelson,HN}.
However the specificity of Markov models
is that the on-site potential $  {\bold H} (x,x) $ is not chosen independently of the off-diagonal matrix elements of Eq. \ref{Hwrates} but is determined in terms of the off-diagonal matrix elements by Eq. \ref{wdiag}
\begin{eqnarray}
{\bold H} (x,x)= - w(x,x)  = - \sum_{x' \ne x}  {\bold H} (x',x)=  \sum_{x' \ne x}   D(x',x) e^{ A(x',x)} 
\label{Hwdiag}
\end{eqnarray}

%%%%%%%%%%%%%%%%%%%%%%%%%%%%%%%%%%%%%%%%%%%%%%%%

\subsection{ Spectral decomposition of the Hamiltonian ${\bold H}$ in the bi-orthogonal basis of right and left eigenvectors }

The spectral decomposition of the non-hermitian Hamiltonian ${\bold H} \ne {\bold H}^{\dagger}$
\begin{eqnarray}
{\bold H} =  \sum_{n=0}^{N-1} E_n \vert r_n \rangle \langle l_n \vert
\label{spectral}
\end{eqnarray}
involves its $N$ eigenvalues $E_n$ that may be complex,  while
 the corresponding right eigenvectors $\vert r_n \rangle $ 
and left eigenvectors $\langle l_n \vert $ satisfy the eigenvalues equations
\begin{eqnarray}
  E_n \vert r_n \rangle && =  {\bold H} \vert r_n \rangle
  \nonumber \\
 E_n \langle l_n \vert && =  \langle l_n \vert{\bold H}  
\label{spectralrl}
\end{eqnarray}
and form a bi-orthogonal basis with the orthonormalization and closure relations
\begin{eqnarray}
\delta_{n,n'} && = \langle l_n \vert  r_{n'} \rangle = \sum_x \langle l_n \vert x \rangle \langle x \vert r_{n'} \rangle
\nonumber \\
{\bold 1}_N && = \sum_{n=0}^{N-1}  \vert r_n \rangle \langle l_n \vert 
\label{orthorl}
\end{eqnarray}

The vanishing eigenvalue $E_0=0$ is associated 
to the left eigenvector unity $l_0(x)=1$ as a consequence of Eq. \ref{wdiag}, 
while the right eigenvector corresponds to the steady state $r_0(x)=p_*(x)$ of Eq. \ref{mastereqsteady}
\begin{eqnarray}
  E_0 && =0
  \nonumber \\
 l_0 (x)&& =1
  \nonumber \\
 r_0 (x)&& =p_*(x)
\label{rlzero}
\end{eqnarray}
The other $(N-1)$ eigenvalues $E_{n=1,..,N-1}$ have strictly positive real parts 
\begin{eqnarray}
{\text {Re}}(E_n)>0 \ \ \ \text{for} \ \ n=1,2,..,N-1
\label{RealEn}
\end{eqnarray}
and govern the relaxation towards the steady state $p_*(x)$ 
of the propagator $p_t(x \vert x_0)$ when one starts at position $x_0$ at time $t=0$ 
\begin{eqnarray}
p_t(x \vert x_0) \equiv \langle x \vert e^{- t {\bold H} } \vert x_0 \rangle 
&& = \sum_{n=0}^{N-1} e^{-t E_n}  \langle x \vert r_n \rangle \langle l_n \vert x_0 \rangle
\nonumber \\
&& = p_*(x) +  \sum_{n=1}^{N-1} e^{-t E_n}  \langle x \vert r_n \rangle \langle l_n \vert x_0 \rangle
\label{propagator}
\end{eqnarray}
So the $(N-1)$ right eigenvectors $r_n(x)=\langle x \vert r_n \rangle$ represent the relaxation modes associated to the excited eigenvalues $E_{n=1,..,N}$.

The spectral decomposition of the propagator of Eq. \ref{propagator}
 is also useful to obtain the relaxation of the average $\overline{ O(x(t))} $
  of any observable $O(x)$ of the configuration $x$
  towards its steady state value $O_*= \sum_x O(x) p_*(x) $.
\begin{eqnarray}
\overline{ O(x(t))} \equiv \sum_x O(x) p_t(x \vert x_0)  
&& =\sum_{n=0}^{N-1} e^{-t E_n} \left( \sum_x O(x) \langle x \vert r_n \rangle \right) \langle l_n \vert x_0 \rangle
\nonumber \\
&& = \left( \sum_x O(x) p_*(x) \right) +  \sum_{n=1}^{N-1} e^{-t E_n} \left( \sum_x O(x) \langle x \vert r_n \rangle \right) \langle l_n \vert x_0 \rangle
\label{observable}
\end{eqnarray}
When the observable $O(x)$ coincides with an excited left eigenvector 
$ \langle l_m \vert x \rangle = l_m^*(x)$, 
the dynamics of Eq. \ref{observable} reduces to a single term
 \begin{eqnarray}
\overline{ l_m^*(x(t))} \equiv \sum_x \langle l_m \vert x \rangle p_t(x \vert x_0)  
&& =\sum_{n=0}^{N-1} e^{-t E_n}   \langle l_m \vert r_n \rangle \langle l_n \vert x_0 \rangle
\nonumber \\
&& =\sum_{n=0}^{N-1} e^{-t E_n}   \delta_{m,n}  \langle l_n \vert x_0 \rangle
=  e^{-t E_m} l_m^*(x_0)
\label{observablelm}
\end{eqnarray}
 So the excited left eigenvector $ \langle l_m \vert x \rangle = l_m^*(x)$ with $m \in \{1,2,..,N-1\}$
 represents a very simple observable
 whose dynamics reduces to the relaxation towards zero 
 with the single exponential governed by the excited eigenvalue $E_m$.

 \subsection{ Discussion   }
 
 In this section, we have recalled how the Markov jump dynamics can be analyzed 
 from the point of view of the dynamics of the probability $p_t(x)$ living on the $N$ configurations $x$. However to better understand the non-equilibrium properties, it is also important
 to analyze the dynamics of the currents $  j_t(x',x)$ defined on the $M$ oriented links,
 as described in the next section.

%%%%%%%%%%%%%%%%%%%%%%%%%%%%%%%%%%%%%%%%%

\section{ Dynamics of the currents $  j_t(x',x)$ defined on the $M$ oriented links }

\label{sec_current}

In this section, we focus on the currents $  j_t(x',x)$  that are defined on the $M$ oriented links existing between the $N$ configurations, in order to better understand their convergence properties towards the steady currents $  j_*(x',x)$ that are non-vanishing whenever the Markov jump process is out-of equilibrium.

\subsection{ Rewriting the master equation as a continuity equation involving the currents defined on the $M$ links   }

On each link between two configurations $(x,x')$ with the parametrization of Eq. \ref{wrates}
for the two transition rates,
it is useful to introduce the antisymmetric current
\begin{eqnarray}
 j_t(x',x) = - j_t(x,'x) && \equiv   w(x',x) p_t(x) - w(x,x') p_t( x' ) 
 \nonumber \\
 && = D(x',x) \left[ e^{A(x',x)} p_t(x) - e^{-A(x',x)}  p_t( x' ) \right]
\label{current}
\end{eqnarray}
Then the master Eq. \ref{mastereq}
can be rewritten as the discrete continuity equation 
\begin{eqnarray}
- \partial_t p_t(x)  =  \sum_{x' \ne x}  j_t(x',x)
\label{continuity}
\end{eqnarray}
where the right handside corresponds to the sum over $x'$ of all the currents $j_t(x',x)$ flowing out of the configuration $x$, i.e. to the discrete divergence at position $x$ of the current.

 %%%%%%%%%%%%%%%%%%%%%%%%%%%%%%%%%%%%%%%%%%

\subsection{ Reminder on the $M$ links between the $N$ configurations and on the $C=M - (N -1 )$ independent cycles  }

The number $M$ of links that connect the $N$ configurations
has for minimal value $M_{min} =N-1$ when the graph is a tree-like structure without any loop,
and for maximal value $M_{max}= \frac{N (N-1)}{2}$ in fully-connected models
when any configuration is connected to the $(N-1)$ other configurations (an example is given in Appendix \ref{app_example}).
In the general case, the difference between the number $M$ of links 
and the minimal value $M_{min} =(N-1)$ needed to connect the $N$ configurations via a spanning tree
(see \cite{luxembourg,phdPolettini,CycleCocycle,gottingen,maes2022,lecomte,Schnakenberg} and references therein)
 \begin{eqnarray}
C \equiv M - (N -1 )
\label{numberofcycles}
\end{eqnarray}
represents the number of independent cycles $\gamma=1,2,..,C$ with the following notations :
\begin{eqnarray} 
&& \text{ a cycle $\gamma$ is a directed self-avoiding closed path of configurations $x^{[\gamma]}(1 \leq l \leq l^{[\gamma]} )$}
\nonumber \\
&& \ \ \ \ \ \ \ \ \ \ \ \ \ \ \ \ \ \text{involving at least $l^{[\gamma]} \geq 3$ distinct configurations}
\nonumber \\
&&  \text{and the same number $l^{[\gamma]} $ 
of oriented links $[x^{[\gamma]}(l) \to x^{[\gamma]}(l+1)]$ with $x^{[\gamma]}(l^{[\gamma]}+1) \equiv x^{[\gamma]}(1)$.}
  \label{cycleGamma}
\end{eqnarray}
 These cycles are essential to characterize the equilibrium or non-equilibrium nature of the steady state as 
 we now recall.
 
 %%%%%%%%%%%%%%%%%%%%%%%%%%%%%%%%%%%%%%%%%%%%%%%%%
 
 %%%%%%%%%%%%%%%%%%%%%%%%%%%%%%%%%%%%%%%%%%
 
 \subsection{ Reminder on the properties of the steady currents $j_* (.,.)$}
 
 \label{subsec_jstar}

The steady current $ j_*(x',x) $ associated to the steady state $p_*(.)$ of Eq. \ref{continuity}
\begin{eqnarray}
 j_*(x',x) = - j_*(x,x') && \equiv   w(x',x) p_*(x) - w(x,x') p_*( x' ) 
  \nonumber \\
 && = D(x,x') \left[ e^{A(x,x')} p_*(x) - e^{-A(x,x')}  p_*( x' ) \right]
\label{currentsteady}
\end{eqnarray}
should be divergenceless, i.e.  for any configuration x, the sum over $x'$ of the steady currents $j_*(x',x) $ out of $x$ should vanish
\begin{eqnarray}
0=  \sum_{x' \ne x}  j_*(x',x)
\label{currentsteadydiv0}
\end{eqnarray}
The vanishing or non-vanishing of the all the link steady currents $ j_*(x',x) $ define the equilibrium or non-equilibrium character of the dynamics as we now recall.

\subsubsection{ Equilibrium steady state $p_*^{eq}(.) $ with vanishing steady currents $j^{eq}_*(.,.)=0$ on all the links }

At equilibrium, the steady current $j_*(.,.)$ of Eq. \ref{currentsteady}
 vanishes on any link
\begin{eqnarray}
0 = j_*^{eq} (x_2,x_1) = w(x_2,x_1) p^{eq}_*(x_1) - w(x_1,x_2) p^{eq}_*(x_2)
= D(x_2,x_1) \left[ e^{A(x_2,x_1)} p^{eq}_*(x_1) - e^{-A(x_1,x_2)}  p^{eq}_*( x_2 ) \right]
\label{Jwhetherzero}
\end{eqnarray}
This vanishing is possible only if the transition rates $w(.,.)$ 
 satisfy
\begin{eqnarray}
1= \frac{  w(x_2,x_1) p^{eq}_*(x_1)}{ w(x_1,x_2) p^{eq}_*(x_2)} = e^{2 A(x_2,x_1) } \frac{   p^{eq}_*(x_1)}{  p^{eq}_*(x_2)} 
\label{DBJ}
\end{eqnarray}
or equivalently only if the antisymmetric function $A(x_2,x_1) $ of Eq. \ref{Avecpot}
corresponds to the discrete gradient 
\begin{eqnarray}
 A(x_2,x_1)  = \frac{1}{2} \left[ \ln \left(  p^{eq}_*(x_2) \right) - \ln \left(  p^{eq}_*(x_1) \right) \right]
\label{Agradient}
\end{eqnarray}
On a tree-like structure without any cycle $C=0$,
the requirement of vanishing divergence of Eq. \ref{currentsteadydiv0}
for the steady currents can be taken into account iteratively starting from the leaves of the tree
to obtain that the steady current should vanish on every link
\begin{eqnarray}
C=0 :   j_*(x',x) =0
\label{CzeroDB}
\end{eqnarray}

On a graph with $C \geq 1 $ independent cycles $\gamma=1,..,C$ introduced around Eq. \ref{cycleGamma},
one needs to check whether the gradient form of Eq. \ref{DBJ} is compatible along each cycle $\gamma$ :
these compatibility conditions can be written either in terms of the transition rates $w(.,.)$ 
in order to recover the famous Kolmogorov criterion for reversibility 
\begin{eqnarray} 
\prod_{l=1}^{l{[\gamma]} } \frac{  w(x^{[\gamma]}(l+1),x^{[\gamma]}(l) ) }{ w(x^{[\gamma]}(l),x^{[\gamma]}(l+1) ) } =1
  \label{KolmogorovcycleGamma}
\end{eqnarray}
or in terms of the antisymmetric function $A(.,.) $ whose total circulation $\Gamma^{[\gamma]}[A(.,.)] $ around each cycle should vanish
\begin{eqnarray} 
\Gamma^{[\gamma]}[A(.,.)]  \equiv \sum_{l=1}^{l{[\gamma]} } A(x^{[\gamma]}(l+1),x^{[\gamma]}(l) ) =0
  \label{KolmogorovcirculationA}
\end{eqnarray}

%%%%%%%%%%%%%%%%%%%%%%%%%%%%%%%%%%%%%%%%%%%%%%%%%%%%%

\subsubsection{ Non-equilibrium steady state $p_*(.) $ with nonvanishing steady currents $j_*(.,.) \ne 0$}

 In all the other cases where the Kolmogorov criterion of Eq. \ref{KolmogorovcycleGamma}  
is not satisfied on the $C$ independent cycles $\gamma=1,..,C$,
or equivalently when the antisymmetric function $A(.,.)$ displays some non-vanishing circulations around cycles 
that prevent its rewriting as a discrete gradient, then
the steady state $p_*(.)$ will be out-of-equilibrium with non-vanishing steady currents 
\begin{eqnarray}
 j_*(.,.)  \ne 0
\label{Jsteadynonzerozero}
\end{eqnarray}

The requirement that the discrete divergence should vanish  (Eq. \ref{currentsteadydiv0})
 yields that the steady current $j_* (x_2,x_1) $ on each oriented link can be written as a as a linear combination of the $C$ cycle-currents $j_*^{Cycle[\gamma]}$ 
that flow around the $C$ independent cycles $\gamma=1,2,..,C$
\begin{eqnarray} 
 j_* (x_2,x_1) 
  = \sum_{\gamma=1}^C j_*^{Cycle[\gamma]} \epsilon^{[\gamma]}(x_2,x_1)
  \label{jsteadycyclesGamma}
\end{eqnarray}
where the coefficient $\epsilon^{[\gamma]}(x_2,x_1)$ takes into account
whether the oriented cycle $\gamma$ contains this oriented link $(x_2 \leftarrow x_1)$ with the same orientation or with the opposite orientation
\begin{eqnarray} 
\epsilon^{[\gamma]}(x_2,x_1)
&& \equiv
  \sum_{ l=1}^{l^{[\gamma]} }  \left(\delta_{x_2,x^{[\gamma]}(l+1)} \delta_{x_1,x^{[\gamma]}(l)} -
  \delta_{x_2,x^{[\gamma]}(l)} \delta_{x_1,x^{[\gamma]}(l+1)}\right)  
\nonumber \\
&&    = \begin{cases}
+1 \text{ if the oriented link $(x_2 \leftarrow x_1)$ appears in the oriented cycle $\gamma$ with the same orientation} 
 \\
-1 \text{ if the oriented link $(x_2 \leftarrow x_1)$ appears in the oriented cycle $\gamma$ with the opposite orientation }  
\\
0 \text{ otherwise }  
\end{cases}
  \label{epsilon}
\end{eqnarray}

In the present paper, we will always assume that the dynamics is out-of-equilibrium with non-vanishing steady current $j_*(.,.) \ne 0$ parametrized by Eq. \ref{jsteadycyclesGamma}
in terms the $C$ cycle-currents associated to the
 $C =M-(N-1) \geq 1$ independent cycles.
Then the steady state $p_*(.)$ should satisfy on each of the $M$ links
\begin{eqnarray}
 w(x_2,x_1) p_*( x_1 ) - w(x_2,x_1) p_*(x_1) 
 = j_*(x_2,x_1) = \sum_{\gamma=1}^C j_*^{Cycle[\gamma]} \epsilon^{[\gamma]}(x_2,x_1)
\label{jjPbasislink}
\end{eqnarray}
This corresponds to a system of $M$ linear equations for the $M=C+(N-1)$ variables
that are the $C$ steady cycle-currents 
$j_*^{Cycle[\gamma]} $ 
and the $(N-1)$ independent coefficients of the normalized steady state $p_*(.)$.

 %%%%%%%%%%%%%%%%%%%%%%%%%%%%%%%%%%%%%%%%%%%%%%%%%%
 
 \subsection{ Factorization of the non-hermitian Hamiltonian $ \bold{H}= {\bold I}^{\dagger}  {\bold J} $
 in terms of two $M \times N$ matrices ${\bold I} $ and ${\bold J} $}

The antisymmetric functions 
\begin{eqnarray}
v(x',x) = - v(x,x')
\label{vanti}
\end{eqnarray}
when the two configurations $x$ and $x'$ at the ends of a link are exchanged
are the discrete analogs of vectors defined in continuous space.
 In order to analyze their properties, it will be useful
to introduce the double-ket notation $\vert _{x_1}^{x_2} \rangle \! \rangle$
 for the space of the $M$ oriented links between two configurations $x_1<x_2$ and to write
\begin{eqnarray}
\langle \! \langle _{x_1}^{x_2} \vert v \rangle \! \rangle \equiv v(x_2,x_1) 
\label{vantiket}
\end{eqnarray}

\subsubsection{ Current matrix ${\bold J}  $ of size $M \times N$}

The current matrix ${\bold J}  $ of size $M \times N$ with the matrix elements
\begin{eqnarray}
\langle \! \langle _{x_1}^{x_2} \vert {\bold J} \vert x \rangle =  w(x_2,x_1) \delta_{x,x_1}  - w(x_1,x_2) \delta_{x,x_2} 
\label{currentOpconfig}
\end{eqnarray}
can be applied to the probability ket $ \vert p_t \rangle$ 
\begin{eqnarray}
\langle \! \langle _{x_1}^{x_2} \vert {\bold J}  \vert p_t \rangle
=  \sum_{x=1}^N \langle \! \langle _{x_1}^{x_2} \vert {\bold J} \vert x \rangle \langle x \vert p_t \rangle
 =  w(x_2,x_1) p_t(x_1) - w(x_1,x_2) p_t(x_2)  
  =   j_t(x_2,x_1)  \equiv \langle \! \langle _{x_1}^{x_2} \vert j_t \rangle \! \rangle
\label{currentlink}
\end{eqnarray}
to reproduce the current $ j_t(x_2,x_1) \equiv \langle \! \langle _{x_1}^{x_2} \vert j_t \rangle \! \rangle
$ of Eq. \ref{current} flowing from $x_1$ towards $x_2$, so that one can write at the matrix level
\begin{eqnarray}
\vert j_t \rangle \! \rangle =  {\bold J}  \vert p_t \rangle
\label{currentlinkop}
\end{eqnarray}
The adjoint matrix ${\bold J}^{\dagger} $ of size $N \times M$ 
acts on a vector $\vert v \rangle  \! \rangle$ to produce the following scalar function of configuration $x$
\begin{eqnarray}
\langle x \vert {\bold J}^{\dagger} \vert v \rangle  \! \rangle 
&& =  \sum_{ \left(_{x_1}^{x_2} \right)}  
\langle x \vert {\bold J}^{\dagger} \vert _{x_1}^{x_2} \rangle  \! \rangle
  \langle \! \langle _{x_1}^{x_2} \vert v \rangle \! \rangle
  = \sum_{ \left(_{x_1}^{x_2} \right)}  \left( w(x_2,x_1) \delta_{x,x_1} -  w(x_1,x_2)\delta_{x,x_2}\right)  
  v(x_2,x_1) 
 \nonumber \\
 &&  = \sum_{ x_2>x}   w(x_2,x) v(x_2,x)
  - \sum_{ x_1<x}   w(x_1,x) v(x,x_1)  
  =  \sum_{ x_2>x}   w(x_2,x) v(x_2,x)
  + \sum_{ x_1<x}   w(x_1,x) v(x_1,x) 
  \nonumber \\
 && = \sum_{x' \ne x}  w(x',x) v(x',x)  
\label{Jdagger}
\end{eqnarray}

%%%%%%%%%%%%%%%%%%%%%%%%%%%%%%%%%%%%%%%%%%

\subsubsection{ Incidence matrix ${\bold I}$ of size $M \times N$}

When all the non-vanishing transition rates $w(x',x) >0 $ are replaced by unity 
\begin{eqnarray}
w(x',x) \to 1
\label{wunity}
\end{eqnarray}
then the current matrix ${\bold J}$ of Eq. \ref{currentOpconfig} 
reduces to the well-known incidence matrix ${\bold I}$ 
that only keeps the information on the geometry of the existing links between configurations
\begin{eqnarray}
\langle \! \langle _{x_1}^{x_2} \vert {\bold I} \vert x \rangle = \delta_{x,x_1} -  \delta_{x,x_2}  
\label{incidence}
\end{eqnarray}

The analog of Eq. \ref{currentlink} 
\begin{eqnarray}
\langle \! \langle _{x_1}^{x_2} \vert {\bold I} \vert f \rangle
&&  = f(x_1) - f(x_2) 
 \equiv - \langle \! \langle _{x_1}^{x_2} \vert \bold{grad} \vert f \rangle
\label{incidencegradient}
\end{eqnarray}
represents the difference of the function $f$ between the two ends of the oriented link,
so the incidence matrix ${\bold I}$ represents the opposite of the discrete gradient 
\begin{eqnarray}
 {\bold I} = -  \bold{grad} 
\label{incidencegradientmatrix}
\end{eqnarray}

The analog of Eq. \ref{Jdagger} involving the adjoint matrix ${\bold I}^{\dagger} $ of size $N \times M$
\begin{eqnarray}
\langle x \vert {\bold I}^{\dagger} \vert v \rangle  \! \rangle   
= \sum_{x' \ne x}   v(x',x)  \equiv \langle x \vert \bold{ div } \vert v \rangle \! \rangle
\label{divergencefromincidence}
\end{eqnarray}
 represents the discrete divergence $\bold{ div } $ of the vector $\vert v \rangle \! \rangle $ at configuration $x$
 \begin{eqnarray}
 {\bold I}^{\dagger} =   \bold{div} 
\label{divergencefromincidencematrix}
\end{eqnarray}

%%%%%%%%%%%%%%%%%%%%%%%%%%%%%%%%%%%%%%%%%%

\subsubsection{ Supersymmetric factorization of the $N \times N$ non-hermitian Hamiltonian into $ \bold{H}= {\bold I}^{\dagger}  {\bold J} $}

   The two matrices $ {\bold J} $ and $ {\bold I} $ of size $M \times N $ introduced above
are useful to factorize the $N \times N $ non-hermitian Hamiltonian $ {\bold H} $ into
\begin{eqnarray}
{\bold H} =   \bold{div}  {\bold J} =   {\bold I}^{\dagger}  {\bold J}
\label{Hfactor}
\end{eqnarray}
This factorization corresponds to the natural splitting of the Euclidean Sch\"odiner Eq. \ref{mastereqschro}
into the following pair of matrix equations 
\begin{eqnarray}
- \partial_t \vert p_t \rangle  && ={\bold I}^{\dagger} \vert j_t \rangle \! \rangle =   \bold{div}  \vert j_t \rangle \! \rangle
\nonumber \\
\vert j_t \rangle \! \rangle && = {\bold J} \vert p_t \rangle
\label{mastereqsplitting}
\end{eqnarray}
that describes the interplay between the probability $\vert p_t \rangle $ living on the $N$ configurations and the current $\vert j_t \rangle \! \rangle $ living on the $M$ oriented links.

%%%%%%%%%%%%%%%%%%%%%%%%%%%%%%%%%%%%%%%%%%

\subsection{ Dynamical properties of the currents defined on the $M$ oriented links }

%%%%%%%%%%%%%%%%%%%%%%%%%%%%%%%%%%%%%%%%%%

\subsubsection{ Dynamics of the current $\vert j_t \rangle \! \rangle  = {\bold J} \vert p_t \rangle $
governed by the supersymmetric partner $
{\hat {\bold H} } \equiv {\bold J}  {\bold I}^{\dagger} $ of size $M \times M$}

A direct consequence of the splitting of Eq. \ref{mastereqsplitting}
 is that the current $\vert j_t \rangle \! \rangle  = {\bold J} \vert p_t \rangle $ 
follows the closed dynamics
\begin{eqnarray}
- \partial_t \vert j_t \rangle \! \rangle && = {\bold J} \bigg( - \partial_t \vert p_t \rangle \bigg) 
= {\bold J} {\bold I}^{\dagger} \vert j_t \rangle \! \rangle
\equiv {\hat {\bold H} } \vert j_t \rangle \! \rangle
\label{mastereqschroj}
\end{eqnarray}
governed by the supersymmetric partner
\begin{eqnarray}
{\hat {\bold H} } \equiv {\bold J}  {\bold I}^{\dagger}  \text{ of size $M \times M$} 
\label{partner}
\end{eqnarray}
of the Hamiltonian ${\bold H}={\bold I}^{\dagger}  {\bold J} $ of dimension $N \times N$.

%%%%%%%%%%%%%%%%%%%%%%%%%%%%%%%%%%%%%%

\subsubsection{ Spectral decomposition of the time-dependent currents $\vert j_t \rangle \! \rangle  $ }

However the size $M \times M$ of the Hamiltonian $ {\hat {\bold H} }$ governing the dynamics of Eq. \ref{mastereqschroj} for the currents $\vert j_t \rangle \! \rangle $ is somewhat misleading.
 Indeed, the currents $\vert j_t \rangle \! \rangle = {\bold J} \vert p_t \rangle $ 
do not live in the full space of dimension $M$ but in a smaller subspace 
since they can be computed from the ket $\vert p_t \rangle $ of dimension $N$
given by the spectral decomposition of the propagator of Eq. \ref{propagator} 
\begin{eqnarray}
 \vert p_t \rangle  && = \sum_{n=0}^{N-1} e^{-t E_n}   \vert r_n \rangle \langle l_n \vert x_0 \rangle
 = \vert p_* \rangle + \sum_{n=1}^{N-1} e^{-t E_n}   \vert r_n \rangle \langle l_n \vert x_0 \rangle
\label{propagatorket}
\end{eqnarray}

So the relaxation of the current 
\begin{eqnarray}
\vert j_t \rangle \! \rangle  = {\bold J} \vert p_t \rangle
&& =  {\bold J}  \vert p_* \rangle + \sum_{n=1}^{N-1} e^{-t E_n} {\bold J}  \vert r_n \rangle \langle l_n \vert x_0 \rangle
\nonumber \\
&& \equiv \vert j_* \rangle \! \rangle + \sum_{n=1}^{N-1} e^{-t E_n}   \vert j_n \rangle \! \rangle \langle l_n \vert x_0 \rangle
\label{jtpropa}
\end{eqnarray}
towards the steady currents $\vert j_* \rangle \! \rangle $ associated to the steady state $\vert p_* \rangle $
\begin{eqnarray}
\vert j_* \rangle \! \rangle && \equiv {\bold J} \vert p_* \rangle
\label{jsteady}
\end{eqnarray}
involves exactly the same $(N-1)$ excited eigenvalues $E_{n=1,..,N-1}$ 
as the Hamiltonian ${\bold H} $.
The corresponding relaxation modes $ \vert j_n \rangle \! \rangle$ 
are the currents associated to the excited right eigenvectors $\vert r_n \rangle$ of  ${\bold H} $
\begin{eqnarray}
\vert j_n \rangle \! \rangle && \equiv {\bold J} \vert r_n \rangle
\label{jn}
\end{eqnarray}
while the eigenvalue Eq. \ref{spectralrl}
for the right eigenvectors $\vert r_n \rangle $
can be then rewritten as
\begin{eqnarray}
  E_n \vert r_n \rangle && = {\bold H} \vert r_n \rangle 
  =  {\bold I}^{\dagger}  {\bold J} \vert r_n \rangle 
  =  {\bold I}^{\dagger}  \vert j_n \rangle \! \rangle 
\label{spectralrnjn}
\end{eqnarray}
The pair of Eqs \ref{jn} and \ref{spectralrnjn}
means that $\vert j_n \rangle \! \rangle $ is a right eigenvector of the partner 
${\hat {\bold H} } \equiv {\bold J}  {\bold I}^{\dagger} $ of Eq. \ref{partner}
associated to the eigenvalue $E_n$
\begin{eqnarray}
  E_n \vert j_n \rangle \! \rangle   = E_n \left({\bold J} \vert r_n \rangle \right)
  = {\bold J}   \left(E_n \vert r_n \rangle \right)
  = {\bold J} {\bold I}^{\dagger}   \vert j_n \rangle \! \rangle 
 ={\hat {\bold H} } \vert j_n \rangle \! \rangle
\label{excitedpartnerright}
\end{eqnarray}
including the case $n=0$ since the non-vanishing steady current $\vert j_{n=0} \rangle \! \rangle = \vert j_* \rangle \! \rangle \ne 0$ is annihilated by $ {\bold I}^{\dagger}  =  \bold{div} $
\begin{eqnarray}
     {\bold I}^{\dagger}   \vert j_* \rangle \! \rangle = 0 ={\hat {\bold H} } \vert j_* \rangle \! \rangle
\label{excitedpartnerrightstar}
\end{eqnarray}

Similarly, the eigenvalue Eq. \ref{spectralrl} for the excited left eigenvector $ \langle l_n \vert  $ 
of ${\bold H} = {\bold I}^{\dagger} {\bold J} $ can be split into
the two matrix equations
\begin{eqnarray}
E_n \langle \! \langle i_n \vert && =   \langle l_n \vert {\bold I}^{\dagger}
\nonumber \\
 \langle l_n \vert  && =   \langle \! \langle  i_n \vert {\bold J} 
\label{spectrallin}
\end{eqnarray}
involving the bra $\langle \! \langle  i_n \vert $ which is a left eigenvector of the partner 
${\hat {\bold H} } \equiv {\bold J}  {\bold I}^{\dagger} $ of Eq. \ref{partner}
associated to the eigenvalue $E_n$
\begin{eqnarray}
E_n \langle \! \langle i_n \vert  
=  \langle l_n \vert {\bold I}^{\dagger}
=  \langle \! \langle i_n \vert {\bold J}   {\bold I}^{\dagger} 
= \langle \! \langle i_n \vert {\hat {\bold H} }
\label{excitedpartnerleft}
\end{eqnarray}
For $n=0$ where the left eigenvector is unity $\langle l_{(n=0)} \vert x \rangle =1$,
the bra $\langle \! \langle i_{(n=0)} \vert  $ satisfies
\begin{eqnarray}
 \langle l_0 \vert  && =   \langle \! \langle  i_0 \vert {\bold J} \ne 0
 \nonumber \\
 0 =  \langle l_0 \vert   {\bold I}^{\dagger}  && =   \langle \! \langle  i_0 \vert {\bold J}  {\bold I}^{\dagger} 
 = \langle \! \langle i_0 \vert {\hat {\bold H} }
\label{spectrallin0}
\end{eqnarray}
The bras $ \langle \! \langle i_{n} \vert$ and the kets $\vert  j_{n'} \rangle \! \rangle $ 
satisfy
the orthonormalization inherited from the orthonormalization of Eq. \ref{orthorl}
concerning the left eigenvectors $\langle l_{n} \vert$ and the right eigenvectors $\vert  r_{n'} \rangle$
\begin{eqnarray}
\langle \! \langle i_{n} \vert  j_{n'} \rangle \! \rangle  =
\langle \! \langle i_{n} \vert  {\bold J} \vert r_{n'} \rangle =  \langle l_{n} \vert  r_{n'} \rangle =  \delta_{n',n}
\label{spectrallinksnorma}
\end{eqnarray}

Note that for the $(N-1)$ excited eigenvalues $E_{n=1,..,N-1} \ne 0$, 
the relations involving the incidence matrix $ {\bold I} = -  \bold{grad}  $ of Eq. \ref{incidencegradientmatrix}
and its adjoint $ {\bold I}^{\dagger} =   \bold{div}  $ of Eq. \ref{divergencefromincidencematrix}
mean that the right eigenvector $r_n(.)$ can be rewritten in terms of the discrete divergence of the $ \vert j_n \rangle \! \rangle$
\begin{eqnarray}
r_n(x)=\langle x   \vert r_n \rangle  =  \frac{1}{E_n}\langle x   \vert {\bold I}^{\dagger}   \vert j_n \rangle \! \rangle 
=  \frac{ \langle x   \vert  \bold{div}   \vert j_n \rangle \! \rangle }{E_n} 
\label{rnasdivergence}
\end{eqnarray}
in constrast to the steady state $ \vert r_{n=0} \rangle = \vert p_* \rangle $ 
associated to the divergenceless steady current 
$\vert j_{n=0} \rangle \! \rangle = \vert j_* \rangle \! \rangle$
\begin{eqnarray}
  \bold{div}   \vert j_* \rangle \! \rangle =0
\label{0divergence}
\end{eqnarray}
while the $\vert i_n  \rangle \! \rangle $ for $n=1,..,N-1$
can be rewritten as the discrete gradient of the left eigenvector $l_n(.) $
\begin{eqnarray}
\langle \! \langle _{x_1}^{x_2} \vert i_n  \rangle \! \rangle&& = 
\frac{1}{E_n^*}  \langle \! \langle _{x_1}^{x_2} \vert {\bold I} \vert l_n \rangle 
= - \frac{1}{E_n^*}\langle \! \langle _{x_1}^{x_2} \vert \bold {grad} \vert l_n \rangle = \frac{  l_n(x_1) - l_n(x_2) }{E_n^*}
\label{lnasgradient}
\end{eqnarray}
in contrast to $ \vert i_{n=0}  \rangle \! \rangle  $ associated to the unity left eigenvector $l_0(x)=1$.

Since the initial current $\vert j_{t=0} \rangle \! \rangle $ associated to the initial condition $\vert x_0 \rangle $ is 
\begin{eqnarray}
\vert j_{t=0} \rangle \! \rangle  = {\bold J} \vert x_0\rangle
\label{jtzero}
\end{eqnarray}
one can rewrite the scalar products $ \langle l_n \vert x_0 \rangle $ appearing in Eq. \ref{jtpropa} as
\begin{eqnarray}
 \langle l_n \vert x_0 \rangle =   \langle \! \langle  i_n \vert {\bold J} \vert x_0 \rangle 
 = \langle \! \langle  i_n \vert  j_{t=0} \rangle \! \rangle
\label{spectrallnexcited}
\end{eqnarray}
including the case $n=0$ with
\begin{eqnarray}
1= l_0(x_0) = \langle l_0 \vert x_0 \rangle =   \langle \! \langle  i_0 \vert {\bold J} \vert x_0 \rangle 
 = \langle \! \langle  i_0 \vert  j_{t=0} \rangle \! \rangle
\label{spectrallnexcitedzero}
\end{eqnarray}
in order to rewrite the spectral decomposition of Eq. \ref{jtpropa} as
\begin{eqnarray}
\vert j_t \rangle \! \rangle  = \vert j_* \rangle \! \rangle
 + \sum_{n=1}^{N-1} e^{-t E_n}   \vert j_n \rangle \! \rangle 
\langle \! \langle  i_n \vert  j_{t=0} \rangle \! \rangle
= \left( \sum_{n=0}^{N-1} e^{-t E_n}   \vert j_n \rangle \! \rangle 
\langle \! \langle  i_n  \vert \right) \vert  j_{t=0} \rangle \! \rangle
\label{jtpropaini}
\end{eqnarray}

%%%%%%%%%%%%%%%%%%%%%%%%%%%%%%%%%%%

 \subsection{ Discussion   }

In summary, the spectral decomposition of Eq. \ref{jtpropaini}
shows that the current $\vert j_t \rangle \! \rangle $ defined on the $M=N-1+C$ oriented links
lives in the subspace of dimension $N$ spanned by the bi-orthogonal basis of
the bras $\langle \! \langle  i_{n=0,..,N-1}  \vert $ and the kets $\vert j_{n=0,1,..,N-1} \rangle \! \rangle $
 that are left and right eigenvectors of the supersymmetric partner ${\hat {\bold H} } $ of Eq. \ref{partner}
 associated to the $N$ eigenvalues $E_{n=0,..,N-1}$ that represent the full spectrum of the $N \times N$ Hamiltonian ${\bold H}  = {\bold I}^{\dagger}  {\bold J}   $.
However since the supersymmetric partner ${\hat {\bold H} } = {\bold J}  {\bold I}^{\dagger} $ 
 is a matrix of size $M \times M$ bigger than $N \times N$,
it is useful to clarify the relations between the spectral properties of the two Hamiltonians 
${\bold H}  = {\bold I}^{\dagger}  {\bold J}   $ and ${\hat {\bold H} } = {\bold J}  {\bold I}^{\dagger} $
via the analysis of the Singular Values Decompositions
of the two rectangular matrices ${\bold I} $ and ${\bold J} $ of size $M \times N$
in the two following sections.

%%%%%%%%%%%%%%%%%%%%%%%%%%%%%%%%%%%%%%%%%%%%%%%%

\section { Singular Value Decomposition of the Incidence matrix ${\bold I} $ of size $M \times N$ }

\label{sec_SVDI}

This section is devoted to the properties of the Singular Value Decomposition 
of the Incidence matrix ${\bold I} $ of Eq. \ref{incidence}
of size $M \times N$ with $M \geq N$.

%%%%%%%%%%%%%%%%%%%%%%%%%%%%%%%%%%%%%%%%%%%%%%%%%%%%%%

\subsection{ SVD of the Incidence matrix ${\bold I} $ involving 
$(N-1)$ positive singular values  $I_{\alpha=1,..,N-1} > 0$ and $ I_{(\alpha=0)} =0$}

The Singular Value Decomposition of the Incidence matrix ${\bold I} $ of size $M \times N$ with $M \geq N$
involves $(N-1)$ strictly positive singular values  $I_{\alpha} > 0$ with $\alpha=1,..,N-1$
\begin{eqnarray}
{\bold I} && =  \sum_{\alpha=1}^{N-1  } I_{\alpha} \vert I^R_{\alpha} \rangle  \! \rangle \langle I^L_{\alpha} \vert
\nonumber \\
{\bold I}^{\dagger} && =  \sum_{\alpha=1}^{N-1  } I_{\alpha} \vert I^L_{\alpha} \rangle \langle \! \langle I^R_{\alpha} \vert
\label{SVDI}
\end{eqnarray}
while the vanishing singular value
\begin{eqnarray}
 I_{(\alpha=0)} =0
\label{Izero}
\end{eqnarray}
is associated to the uniform normalized eigenvector
\begin{eqnarray}
\langle x \vert I^L_{(\alpha=0)} \rangle = \frac{1}{\sqrt N} 
\label{Uniform}
\end{eqnarray}
that is annihilated by the matrix ${\bold I} $ that represents the opposite of the discrete gradient (Eq. \ref{incidencegradient})
\begin{eqnarray}
{\bold I} \vert I^L_{(\alpha=0)} \rangle =  -  \bold{grad} \vert I^L_{(\alpha=0)} \rangle = 0
\label{UniformAnnihilated}
\end{eqnarray}

The $N$ left singular kets $\vert I^L_{\alpha=0,1..,N-1} \rangle $ form an orthonormal basis of the space of the $N$ configurations
 \begin{eqnarray}
\delta_{\alpha,\alpha'} && = \langle I^L_{\alpha} \vert I^L_{\alpha'} \rangle 
= \sum_x \langle I^L_{\alpha} \vert x \rangle \langle x \vert I^L_{\alpha'} \rangle
\nonumber \\
{\bold 1}_N && = \sum_{\alpha=0}^{N-1  }  \vert I^L_{\alpha} \rangle \langle I^L_{\alpha} \vert
= \sum_{\alpha=1}^{N-1  }  \vert I^L_{\alpha} \rangle \langle I^L_{\alpha} \vert 
+   \vert I^L_{(\alpha=0)} \rangle \langle I^L_{(\alpha=0)} \vert
\label{orthoIL}
\end{eqnarray}
while the $(N-1)$ right singular kets $\vert I^R_{\alpha=1,..,N-1} \rangle \! \rangle $ 
that are associated to the $(N-1)$ strictly positive singular values  $I_{\alpha=1,..,N-1} > 0$ in Eq. \ref{SVDI}
should be supplemented 
by $M-(N-1)=C$ other kets 
$\vert  I^R_{\alpha=0,-1,-2..,-(C-1)} \rangle \! \rangle $ in order to obtain an orthonormal basis of the space of the $M$ oriented links
  \begin{eqnarray}
\delta_{\alpha,\alpha'} && =  \langle \!\langle I^R_{\alpha} \vert I^R_{\alpha'} \rangle \! \rangle
=  \sum_{ \left(_{x_1}^{x_2} \right)}    \langle \!\langle I^R_{\alpha} \vert _{x_1}^{x_2} \rangle \! \rangle
\langle \! \langle _{x_1}^{x_2} \vert I^R_{\alpha'} \rangle \! \rangle
\nonumber \\
{\bold 1}_M && = \sum_{\alpha=-(C-1)}^{N-1  }  \vert I^R_{\alpha} \rangle \! \rangle \langle \! \langle I^R_{\alpha} \vert
= \sum_{\alpha=1}^{N-1  }  \vert I^R_{\alpha} \rangle \! \rangle \langle \! \langle I^R_{\alpha} \vert
+ \sum_{\alpha=-(C-1)}^{0  }  \vert I^R_{\alpha} \rangle \! \rangle \langle \! \langle I^R_{\alpha} \vert
\label{orthoIR}
\end{eqnarray}

%%%%%%%%%%%%%%%%%%%%%%%%%%%%%%%%%%%%%%%%%%%%%%%%%%%%%%

\subsection{ Relation with the spectral decomposition of the discrete Laplacian in the space of the $N$ configurations }

Since the incidence matrix ${\bold I}$ represents the opposite of the discrete gradient (Eq. \ref{incidencegradientmatrix})
and since the adjoint matrix ${\bold I}^{\dagger} $ represents the discrete divergence (Eq. \ref{divergencefromincidencematrix}),
the supersymmetric matrix ${\bold I}^{\dagger} {\bold I} $ of size $N \times N$ 
corresponds to the opposite of the discrete Laplacian ${\bold \Delta} $ in the space of the $N$ configurations
\begin{eqnarray}
{\bold I}^{\dagger}{\bold I} && = {\bold {div}}(- {\bold {grad}} ) =- {\bold \Delta}
\label{Laplacian}
\end{eqnarray}
as can be checked via the evaluation of the matrix elements using Eq. \ref{incidence}
\begin{eqnarray}
\langle x \vert {\bold I}^{\dagger} {\bold I}  \vert x' \rangle && 
= \sum_{ \left(_{x_1}^{x_2} \right)} \langle x' \vert {\bold I} \vert _{x_1}^{x_2} \rangle \! \rangle
 \langle x' \vert {\bold I} \vert _{x_1}^{x_2} \rangle  \! \rangle
= \sum_{ \left(_{x_1}^{x_2} \right)}  \big(  \delta_{x,x_1} -  \delta_{x,x_2} \big) 
\big(  \delta_{x',x_1} -  \delta_{x',x_2} \big) 
\nonumber \\
&& = \sum_{ \left(_{x_1}^{x_2} \right)}  
\big(  \delta_{x,x_1}  \delta_{x',x_1} +\delta_{x,x_2} \delta_{x',x_2} 
 - \delta_{x,x_2} \delta_{x',x_1}
- \delta_{x,x_1}  \delta_{x',x_2}  \big) 
\nonumber \\
&& 
= \begin{cases}
\text{ $z(x)$ if $x=x'$} 
 \\
-1 \text{  if $x$ and $x'$ are the two ends of a link}  
\\
0 \text{ otherwise }  
\end{cases}
\equiv - \langle x \vert {\bold \Delta} \vert x' \rangle
\label{incidenceidagger}
\end{eqnarray}
where $z(x)$ is the number of links connected to the configuration $x$.

On the other hand, the evaluation of the supersymmetric matrix ${\bold I}^{\dagger} {\bold I} $ via the Singular Value  Decompositions of Eq. \ref{SVDI} for the incidence matrix ${\bold I}$ and its adjoint ${\bold I}^{\dagger} $
\begin{eqnarray}
- {\bold \Delta} = {\bold I}^{\dagger}{\bold I} && = \left( \sum_{\alpha=1}^{N-1  } I_{\alpha} \vert I^L_{\alpha} \rangle \langle \! \langle I^R_{\alpha} \vert\right) 
\left(  \sum_{\alpha'=1}^{N-1  } I_{\alpha'} \vert I^R_{\alpha'} \rangle  \! \rangle \langle I^L_{\alpha'} \vert\right) 
=  \sum_{\alpha=1}^{N -1 } I_{\alpha}^2 \vert I^L_{\alpha} \rangle \langle I^L_{\alpha} \vert
\label{SVDdaggerII}
\end{eqnarray}
gives its spectral decomposition in terms of the $(N-1)$ strictly positive eigenvalues $I_{\alpha=1,..,N-1}^2>0$
with their associated orthonormalized eigenvectors $\vert I^L_{\alpha=1,..,N-1} \rangle $,
while the vanishing eigenvalue $ I^2_{(\alpha=0)} =0 $ of Eq. \ref{Izero}
is associated to the uniform normalized eigenvector $\vert I^L_{(\alpha=0)} \rangle $ of Eq. \ref{Uniform}.

In particular, whenever
the orthonormalized basis of eigenvectors $\vert I^L_{\alpha} \rangle $ of the opposite Laplacian 
$(- {\bold \Delta} ) $ of Eq. \ref{SVDdaggerII}
is known, this is the the orthonormalized basis of left singular vectors $\vert I^L_{\alpha} \rangle $
that appear in the SVD of the incidence matrix ${\bold I}$ of Eq. \ref{SVDI},
while the corresponding singular values $I_{\alpha} $ of the incidence matrix ${\bold I}$ 
are given by the square-roots of the eigenvalues $I_{\alpha}^2 $.
The corresponding $(N-1)$ right singular vectors $ \vert I^R_{\alpha=1,..,N-1} \rangle$ of Eq. \ref{SVDI}
can be then obtained via the application of the incidence matrix ${\bold I} $ on the left singular vectors $\vert I^L_{\alpha} \rangle $
\begin{eqnarray}
{\bold I} \vert I^L_{\alpha} \rangle  && = \left(  \sum_{\alpha'=1}^{N -1  } I_{\alpha'} \vert I^R_{\alpha'} \rangle  \! \rangle \langle I^L_{\alpha'} \vert \right) \vert I^L_{\alpha} \rangle 
= I_{\alpha} \vert I^R_{\alpha} \rangle  \! \rangle
\label{RfromL}
\end{eqnarray}
The further application of the adjoint matrix ${\bold I}^{\dagger}  $ to these right singular vectors $ \vert I^R_{\alpha=1,..,N-1} \rangle \! \rangle$
\begin{eqnarray}
{\bold I}^{\dagger} \vert I^R_{\alpha} \rangle  \! \rangle  
&& = \left(  \sum_{\alpha'=1}^{N -1  } I_{\alpha'} \vert I^L_{\alpha'} \rangle \langle \! \langle I^R_{\alpha'} \vert  \right) 
\vert I^R_{\alpha} \rangle  \! \rangle
= I_{\alpha} \vert I^L_{\alpha} \rangle 
\label{LfromR}
\end{eqnarray}
then reproduce the left singular vectors $\vert I^L_{\alpha} \rangle $.

%%%%%%%%%%%%%%%%%%%%%%%%%%%%%%%%%%%%%%%%%%%%%%%%%%%%%

\subsection{ Relation with the spectral decomposition of the supersymmetric partner ${\bold I}  {\bold I}^{\dagger} $ 
of size $M \times M$}

The matrix elements of the supersymmetric partner ${\bold I} {\bold I}^{\dagger} $ of size $M \times M$
read using Eq. \ref{incidence}
\begin{eqnarray}
&& \langle _{x_1}^{x_2} \vert  {\bold I}  {\bold I}^{\dagger}\vert _{x_1'}^{x_2'} \rangle  
=  \sum_{ x} \langle _{x_1}^{x_2} \vert {\bold I} \vert  x \rangle \langle _{x_1'}^{x_2'} \vert {\bold I} \vert  x \rangle
= \sum_{ x} \big(  \delta_{x,x_1} -  \delta_{x,x_2} \big) 
\big(  \delta_{x,x_1'} -  \delta_{x,x_2'} \big)
\nonumber \\
&& =  \delta_{x_1,x_1'} + \delta_{x_2,x_2'} -  \delta_{x_1,x_2'}  - \delta_{x_1',x_2} 
\nonumber \\
&&= \begin{cases}
\text{ 2 if the two oriented links coincide, i.e. $x_1=x_1'$ and $x_2=x_2'$} 
 \\
\text{ 1 if the two oriented links are different but share the same starting-point $x_1=x_1'$ or the same end-point $x_2=x_2' \ \ \ \ $}  
 \\
\text{ -1 if the starting-point of one oriented link coincide with the end-point of the other link, i.e. $x_2=x_1'$ or $x_2'=x_1$}  
\end{cases}
\label{idaggerincidence}
\end{eqnarray}

On the other hand, the evaluation of the supersymmetric matrix ${\bold I} {\bold I}^{\dagger} $ of size $M \times M $
 via the Singular Value  Decompositions of Eq. \ref{SVDI} for the incidence matrix ${\bold I}$ and its adjoint ${\bold I}^{\dagger} $
\begin{eqnarray}
{\bold I} {\bold I}^{\dagger} && =  \sum_{\alpha=1}^{N -1 } I_{\alpha}^2 
\vert I^R_{\alpha} \rangle  \! \rangle \langle \! \langle I^R_{\alpha} \vert 
\label{SVDIdaggerI}
\end{eqnarray}
gives its spectral decomposition that 
involves the same $(N -1)$ strictly positive eigenvalues $I_{\alpha=1,..,N-1}^2 >0$ 
as the opposite-Laplacian of Eq. \ref{SVDdaggerII},
while the corresponding eigenvectors $ \vert I^R_{\alpha=1,..,N-1} \rangle  \! \rangle $ are 
related to the eigenvectors $\vert I^L_{\alpha=1,..,N-1} \rangle $ of the opposite-Laplacian via Eq. \ref{RfromL}.

So here the vanishing eigenvalue $I^2_0=0 $ 
is degenerate and associated to the subspace of dimension $M-(N-1)=C$
with the orthonormalized basis $\vert I^R_{\alpha=0,-1,..,-(C-1)} \rangle  \! \rangle $.

%%%%%%%%%%%%%%%%%%%%%%%%%%%%%%%%%%%%%%%%%%%%

\subsection{ Relation with the discrete Helmholtz decomposition for an arbitrary vector $\vert v \rangle  \! \rangle $ } 

\label{subsec_discreteHelm}

An arbitrary vector $\vert v \rangle  \! \rangle $ in the space of dimension $M$ 
can be decomposed with respect to the orthonormalized basis $\vert I^R_{\alpha} \rangle  \! \rangle $ of Eq. \ref{orthoIR}
in terms of its $M$ coefficients 
  \begin{eqnarray}
v_{\alpha} \equiv  \langle \! \langle I^R_{\alpha} \vert v \rangle  \! \rangle \ \ \text{for } \alpha=-(C-1),...,-1,0,+1,..,N-1
\label{coefsvalpha}
\end{eqnarray}
It is useful to separate these $M=(N-1)+C$ terms into two orthogonal contributions 
of dimensions $(N-1)$ and $C$ respectively
  \begin{eqnarray}
\vert v \rangle  \! \rangle && = \sum_{\alpha=-(C-1)}^{N-1} v_{\alpha}\vert I^R_{\alpha} \rangle  \! \rangle\equiv \vert v^{[I_.>0]} \rangle  \! \rangle + \vert v^{[I_0=0]} \rangle  \! \rangle
\nonumber \\
\vert v^{[I_.>0]} \rangle  \! \rangle&& \equiv  \sum_{\alpha=1}^{N-1} v_{\alpha}\vert I^R_{\alpha} \rangle  \! \rangle  
\nonumber \\
\vert v^{[I_0=0]} \rangle  \! \rangle && \equiv \sum_{\alpha=-(C-1)}^{0} v_{\alpha}\vert I^R_{\alpha} \rangle  \! \rangle 
\label{discreteHelmhotz}
\end{eqnarray}
with the following properties.

%%%%%%%%%%%%%%%%%%%%%%%%%%%%%%%%%%%%%%%%%%

\subsubsection{ The component $\vert v^{[I_0=0]} \rangle  \! \rangle $ of dimension $C$ associated to the 
vanishing singular value $I_0=0 $ } 

\label{subseccycle}

The vanishing singular value $I_0=0 $ 
is degenerate and associated to the subspace of dimension $C = M - (N -1 )$ 
that is annihilated by the adjoint operator ${\bold I}^{\dagger}=\bold {div}$ 
of Eq. \ref{divergencefromincidencematrix}
corresponding to the discrete divergence.
As a consequence, the component $\vert v^{[I_0=0]} \rangle  \! \rangle $ of Eq. \ref{discreteHelmhotz}
can be characterized by its vanishing divergence
  \begin{eqnarray} 
0 = {\bold I}^{\dagger} \vert v^{[I_0=0]} \rangle  \! \rangle =\bold {div}  \vert v^{[I_0=0]} \rangle  \! \rangle
  \label{annihIdagger}
\end{eqnarray}
As already discussed around Eq. \ref{jsteadycyclesGamma} on the special case of the steady current $\vert j_* \rangle  \! \rangle$ whose divergence vanishes $\bold {div}  \vert j_* \rangle  \! \rangle =0$,
the divergenceless component  $\vert v^{[I_0=0]} \rangle  \! \rangle$ 
can be similarly parametrized by
$C$ cycle-components flowing around the $C$ independent cycles $\gamma=1,2,..,C$.

%%%%%%%%%%%%%%%%%%%%%%%%%%%%%%%%%%%%%%%%%%%%%%%%%%%%

\subsubsection{ The component $\vert v^{[I_.>0]} \rangle  \! \rangle $ of dimension $(N -1 )$ associated to the 
strictly positive singular values $I_{\alpha=1,2,..,N-1}>0 $  } 

\label{subsecgrad}

The component $\vert v^{[I_.>0]} \rangle  \! \rangle $ of Eq. \ref{discreteHelmhotz}
associated to the 
$(N-1)$ strictly positive singular values $I_{\alpha=1,2,..,N-1}>0 $
can be rewritten using Eq. \ref{RfromL}
\begin{eqnarray}
\vert v^{[I_.>0]} \rangle  \! \rangle  = \sum_{\alpha=1}^{N-1} v_{\alpha}\vert I^R_{\alpha} \rangle  \! \rangle
= \sum_{\alpha=1}^{N-1} \frac{ v_{\alpha} }{ I_{\alpha} } {\bold I} \vert I^L_{\alpha} \rangle  
 = {\bold I} \left( \sum_{\alpha=1}^{N-1} \frac{ v_{\alpha} }{ I_{\alpha} } \vert I^L_{\alpha} \rangle \right) 
\equiv - {\bold {grad} } \vert g^{[I_.>0]} \rangle 
\label{vGradbasisRfromL}
\end{eqnarray}
as the opposite discrete gradient ${\bold I}  
\equiv - {\bold {grad} } $ of Eq. \ref{incidencegradientmatrix}
applied to the following ket $\vert g^{[I_.>0]} \rangle $ that lives in the space of the $N$ configurations
\begin{eqnarray}
\vert g^{[I_.>0]} \rangle   \equiv \sum_{\alpha=1}^{N-1} \frac{ v_{\alpha} }{ I_{\alpha} } \vert I^L_{\alpha} \rangle 
\label{uvgrad}
\end{eqnarray}
and that belongs to the subspace spanned by the $(N-1)$ left singular vectors $ \vert I^L_{\alpha=1,2,..,N-1} \rangle $  
orthogonal to the constant ket $ \vert I_{(\alpha=0)} \rangle$ of Eq. \ref{Uniform}.
As a consequence, the circulation of $\vert v^{[I_.>0]} \rangle  \! \rangle  =  - {\bold {grad} } \vert g^{[I_.>0]} \rangle $
of Eq. \ref{vGradbasisRfromL} around any of the $C$ independent cycle $\gamma=1,2,..,C$ vanishes
  \begin{eqnarray} 
\Gamma^{[\gamma]}[ v^{[I_.>0]} ]  \equiv 
\sum_{l=1}^{l{[\gamma]} } \langle \! \langle ^{x^{[\gamma]}(l+1)}_{x^{[\gamma]}(l)} \vert v^{[I_.>0]} \rangle  \! \rangle =0
  \label{circulationvgrad}
\end{eqnarray}

The application of the discrete divergence ${\bold I}^{\dagger}=\bold {div}$ of Eq. \ref{divergencefromincidencematrix}
to the component $\vert v^{[I_.>0]} \rangle  \! \rangle  =  - {\bold {grad} } \vert g^{[I_.>0]} \rangle $
of Eq. \ref{vGradbasisRfromL}
corresponds to the application of the opposite Laplacian of Eq. \ref{SVDdaggerII} 
to the ket $\vert g^{[I_.>0]} \rangle $ of Eq. \ref{uvgrad}
\begin{eqnarray}
{\bold I}^{\dagger} \vert v^{[I_.>0]} \rangle  \! \rangle && = {\bold I}^{\dagger}{\bold I} \vert g^{[I_.>0]} \rangle
= - {\bold \Delta}  \vert g^{[I_.>0]} \rangle
\label{helmLaplacian}
\end{eqnarray}
with the following decomposition in the orthonormalized basis $\vert I^L_{\alpha} \rangle $ of the Laplacian
associated to the eigenvalues $I_{\alpha}^2 $
\begin{eqnarray}
{\bold I}^{\dagger} \vert v^{[I_.>0]} \rangle  \! \rangle && =  - {\bold \Delta} \left(  \sum_{\alpha=1}^{N-1} \frac{ v_{\alpha} }{ I_{\alpha} } \vert I^L_{\alpha} \rangle \right)
= \sum_{\alpha=1}^{N-1}  v_{\alpha} I_{\alpha}  \vert I^L_{\alpha} \rangle 
\label{helmLaplaciianalpha}
\end{eqnarray}

%%%%%%%%%%%%%%%%%%%%%%%%%%%%%%%%%%%%%%%%%%%%%%%%%%%%

\subsubsection{ Conclusion on the discrete Helmholtz decomposition associated to the SVD of the incidence matrix ${\bold I}$  } 

In conclusion, for an arbitrary vector $\vert v \rangle  \! \rangle $ of the space of $M$ links,
the decomposition of Eq. \ref{discreteHelmhotz}
into its gradient component $\vert v^{[I_.>0]} \rangle  \! \rangle  = - {\bold {grad} } \vert g^{[I_.>0]} \rangle$  
of Eq. \ref{vGradbasisRfromL} of dimension $(N-1)$
and into its divergenceless component $\vert v^{[I_0=0]} \rangle  \! \rangle $ of Eq. \ref{annihIdagger} of dimension $C$
corresponds to the discrete Helmholtz decomposition,
with the two important properties:

(1) The application of the discrete divergence ${\bold I}^{\dagger}=\bold {div}$ on 
the vector $\vert v \rangle  \! \rangle $ 
only involves the divergence of the gradient component $\vert v^{[I_.>0]} \rangle  \! \rangle  = - {\bold {grad} } \vert g^{[I_.>0]} \rangle$ 
  \begin{eqnarray}
\bold {div}\vert v \rangle  \! \rangle = {\bold I}^{\dagger} \vert v \rangle  \! \rangle && 
= {\bold I}^{\dagger} \vert v^{[I_.>0]} \rangle  \! \rangle + 0
= - {\bold \Delta } \vert g^{[I_.>0]} \rangle
\label{divergencediscreteHelmhotz}
\end{eqnarray}
and reduces to the application of the opposite Laplacian to the ket $\vert g^{[I_.>0]} \rangle $.

(2)The circulation of the vector $\vert v \rangle  \! \rangle $ along the any of the $C$ independent closed cycles $\gamma=1,..,C$ 
only involves the circulation of the divergenceless component $\vert v^{[I_0=0]} \rangle  \! \rangle$
   \begin{eqnarray} 
\Gamma^{[\gamma]}[ v ]  \equiv 
\sum_{l=1}^{l{[\gamma]} } \langle \! \langle ^{x^{[\gamma]}(l+1)}_{x^{[\gamma]}(l)} \vert v \rangle  \! \rangle
= 0 + \Gamma^{[\gamma]}[ v^{[I_0=0]} ] 
  \label{circulationvtot}
\end{eqnarray}

%%%%%%%%%%%%%%%%%%%%%%%%%%%%%%%%%%%%%%%%%%%%%%%

\subsubsection{ Application to the definition of the pseudo-inverse ${\bold I}^{pseudo[-1]} $ of the incidence matrix ${\bold I} $}

\label{subsec_pseudoInversionI}

The linear system involving the rectangular incidence matrix ${\bold I}$ 
\begin{eqnarray}
{\bold I} \vert g \rangle =\vert v \rangle \! \rangle
\label{linearsystemI}
\end{eqnarray}
for the unknown ket $\vert g \rangle $ of dimension $N$ 
when the vector $\vert v \rangle \! \rangle$ of dimension $M$ is given
can be analyzed by plugging the SVD of Eq. \ref{SVDI}
for the incidence matrix into Eq. \ref{linearsystemI}
\begin{eqnarray}
\sum_{\alpha=1}^{N-1  } I_{\alpha} \vert I^R_{\alpha} \rangle  \! \rangle \langle I^L_{\alpha} \vert  g \rangle =\vert v \rangle \! \rangle
\label{linearsystemISDV}
\end{eqnarray}
and via the Helmholtz decomposition of Eq. \ref{discreteHelmhotz}
for the arbitrary vector $\vert v \rangle  \! \rangle $ in the space of dimension $M$
with the following discussion :

(i) The component  $\vert v^{[I_0=0]} \rangle  \! \rangle$ 
spanned by the kets $\vert I^R_{\alpha=0,-1,..,-(C-1)} \rangle  \! \rangle $ that do not appear on the left hand side
of Eq. \ref{linearsystemISDV}
should vanish
\begin{eqnarray}
\vert v^{[I_0=0]} \rangle  \! \rangle && \equiv \sum_{\alpha=-(C-1)}^{0} v_{\alpha}\vert I^R_{\alpha} \rangle  \! \rangle =0
\label{linearsystemISDVzero}
\end{eqnarray}
i.e. the $C$ coefficients $ v_{\alpha=0,-1,..,-(C-1)} $ should vanish
\begin{eqnarray}
v_{\alpha} \equiv \langle \! \langle I^R_{\alpha} \vert v \rangle  \! \rangle =0  \ \ \text{ for } \alpha=0,-1,..,-(C-1)
\label{linearsystemISDVzerocoefs}
\end{eqnarray}

(ii) The identification of the coefficients of the $(N-1)$ eigenvectors $\vert I^R_{\alpha=1,2,..,N-1} \rangle  \! \rangle $ in Eq. \ref{linearsystemISDV}
yields that the coefficients of $\vert  g \rangle $ in the basis of the left singular vectors $ \vert I^L_{\alpha} \rangle $
are given by
  \begin{eqnarray}
   \langle I^L_{\alpha} \vert  g \rangle = \frac{ v_{\alpha} }{ I_{\alpha} } 
   \ \ \text{ for } \alpha=1,2,..,N-1
\label{linearsystemISDVnonzero}
\end{eqnarray}
so that the solution for the ket $\vert  g \rangle $
  \begin{eqnarray}
\vert  g \rangle  = \sum_{\alpha=1}^{N-1}  \vert I^L_{\alpha} \rangle \langle I^L_{\alpha} \vert  g \rangle 
=\sum_{\alpha=1}^{N-1}  \vert I^L_{\alpha} \rangle  \frac{ \langle \! \langle I^R_{\alpha} \vert v \rangle  \! \rangle }{ I_{\alpha} } 
= \left( \sum_{\alpha=1}^{N-1}    \frac{ \vert I^L_{\alpha} \rangle \langle \! \langle I^R_{\alpha} \vert  }{ I_{\alpha} } \right) \vert v \rangle  \! \rangle
\equiv {\bold I}^{pseudo[-1]} \vert v \rangle  \! \rangle
\label{linearsystemISDVsolution}
\end{eqnarray}
corresponds to the application to the vector $\vert v \rangle  \! \rangle $
of the pseudo-inverse ${\bold I}^{pseudo[-1]} $ of the incidence matrix ${\bold I} $ with the SVD of Eq. \ref{SVDI}
\begin{eqnarray}
{\bold I}^{pseudo[-1]} \equiv \sum_{\alpha=1}^{N-1}    \frac{ \vert I^L_{\alpha} \rangle \langle \! \langle I^R_{\alpha} \vert  }{ I_{\alpha} } 
\label{SVDIpseudo}
\end{eqnarray}

%%%%%%%%%%%%%%%%%%%%%%%%%%%%%%%%%%%%

 \subsection{ Discussion   }

In summary, the Singular Value Decomposition of the incidence matrix ${\bold I} $ of size $M \times N$ is directly related to the discrete Helmholtz decomposition of an arbitrary vector $\vert v \rangle  \! \rangle $ of dimension $M$ into its gradient part of dimension $(N-1)$ and its divergenceless 
part of dimension $C$, related to the $C$ independent cycles, that were previously introduced to parametrize the divergenceless steady current $j_*(.,.)$ as recalled around Eq. \ref{jsteadycyclesGamma}. However, in order to understand the subspace of dimension $N$
where the time-dependent current $j_t(.,.)$ live, one needs to consider 
the Singular Value Decomposition of the current matrix ${\bold J} $, as discussed in the next section.

%%%%%%%%%%%%%%%%%%%%%%%%%%%%%%%%%%%%%%%%%%%%%%%%

\section { Singular Value Decomposition of the current matrix ${\bold J} $ of size $M \times N$ }

\label{sec_SVDJ}

This section is devoted to the properties of the Singular Value Decomposition
 of the current matrix ${\bold J} $ of Eq. \ref{currentOpconfig}
of size $M \times N$ with $M \geq N$.
Since the current matrix ${\bold J} $ of Eq. \ref{currentOpconfig}
can be considered as a deformation by the transition rates $w(.,.)$ of the incidence matrix ${\bold I}  $ of Eq. \ref{incidence},
whose SVD was discussed in detail in the previous section \ref{sec_SVDI},
many properties are very similar, so we will mainly stress the important differences.

%%%%%%%%%%%%%%%%%%%%%%%%%%%%%%%%

\subsection{ SVD of the current matrix ${\bold J} $ involving $N$ strictly positive singular values  
$ \lambda_{\beta=1,..,N}  > 0$ }

For a non-equilibrium steady state with non-vanishing steady currents (Eq. \ref{Jsteadynonzerozero}),
 the Singular Value Decomposition of the current matrix ${\bold J} $ of size $M \times N$
 involves $N$ strictly positive singular values  $\lambda_{\beta=1,..,N} >0 $ 
\begin{eqnarray}
{\bold J} && =  \sum_{\beta=1}^{N  } \lambda_{\beta} 
\vert \lambda^R_{\beta} \rangle  \! \rangle \langle \lambda^L_{\beta} \vert
\nonumber \\
{\bold J}^{\dagger} && =  \sum_{\beta=1}^{N  } \lambda_{\beta} 
\vert \lambda^L_{\beta} \rangle \langle \! \langle \lambda^R_{\beta} \vert
\label{SVDJ}
\end{eqnarray}
where the $N$ right singular vectors $\vert \lambda^R_{\beta=1,..,N} \rangle  \! \rangle$ 
should be supplemented 
by $M-N=(C-1)$ other vectors 
$\vert \lambda^R_{\beta=0,-1..,-(C-2)} \rangle  \! \rangle $ 
in order to obtain an orthonormal basis of the space of the $M$ oriented links
  \begin{eqnarray}
\delta_{\beta,\beta'} && =  \langle \!\langle \lambda^R_{\beta} \vert \lambda^R_{\beta'} \rangle \! \rangle
=  \sum_{ \left(_{x_1}^{x_2} \right)}    \langle \!\langle \lambda^R_{\beta} \vert _{x_1}^{x_2} \rangle \! \rangle
\langle \! \langle _{x_1}^{x_2} \vert \lambda^R_{\beta'} \rangle \! \rangle
\nonumber \\
{\bold 1}_M && = \sum_{\beta=-(C-2)}^{N  }  \vert \lambda^R_{\beta} \rangle \! \rangle \langle \! \langle \lambda^R_{\beta} \vert
= \sum_{\beta=1}^{N  }  \vert \lambda^R_{\beta} \rangle \! \rangle \langle \! \langle \lambda^R_{\beta} \vert
+ \sum_{\beta=-(C-2)}^{0  }  \vert \lambda^R_{\beta} \rangle \! \rangle \langle \! \langle \lambda^R_{\beta} \vert
\label{orthoJR}
\end{eqnarray}
while the $N$ left singular vectors $\vert \lambda^L_{\beta=1,2..,N} \rangle $ form an orthonormal basis of the space of the $N$ configurations
 \begin{eqnarray}
\delta_{\beta,\beta'} && = \langle \lambda^L_{\beta} \vert \lambda^L_{\beta'} \rangle 
= \sum_x \langle \lambda^L_{\beta} \vert x \rangle \langle x \vert \lambda^L_{\beta'} \rangle
\nonumber \\
{\bold 1}_N && = \sum_{\beta=1}^{N  }  \vert \lambda^L_{\beta} \rangle \langle \lambda^L_{\beta} \vert
\label{orthoJL}
\end{eqnarray}
that can be found 
from the diagonalization of the $N \times N$ supersymmetric matrix   
\begin{eqnarray}
 {\bold J}^{\dagger}{\bold J} && 
=  \sum_{\beta=1}^{N  } \lambda_{\beta}^2 \vert \lambda^L_{\beta} \rangle \langle \lambda^L_{\beta} \vert
\label{SVDdaggerJJ}
\end{eqnarray}
whose matrix elements read
\begin{eqnarray}
&& \langle x \vert {\bold J}^{\dagger} {\bold J}  \vert x' \rangle  
= \sum_{ \left(_{x_1}^{x_2} \right)} \langle x \vert {\bold J} \vert _{x_1}^{x_2} \rangle \! \rangle
 \langle x' \vert {\bold J} \vert _{x_1}^{x_2} \rangle  \! \rangle
= \sum_{ \left(_{x_1}^{x_2} \right)}  \big( w(x_2,x_1) \delta_{x,x_1} - w(x_1,x_2) \delta_{x,x_2} \big) 
\big( w(x_2,x_1) \delta_{x',x_1} - w(x_1,x_2) \delta_{x',x_2} \big) 
\nonumber \\
&& = \sum_{ \left(_{x_1}^{x_2} \right)}  
\big( w^2(x_2,x_1)\delta_{x,x_1}  \delta_{x',x_1}
+w^2 (x_1,x_2) \delta_{x,x_2} \delta_{x',x_2} 
 - w(x_2,x_1)  w(x_1,x_2)\delta_{x,x_2} \delta_{x',x_1}
- w(x_2,x_1)  w(x_1,x_2)\delta_{x,x_1}  \delta_{x',x_2}  \big) 
\nonumber \\
&& 
= \begin{cases}
\displaystyle \sum_{ x'' \ne x}  w^2(x'',x) \ \ \ \ \ \ \ \ \ \ \ \ \ \ \ \ \ \ \ \ \ \ \ \ \ \ \ \  \text{   if }  x=x'  
 \\
- w(x,x') w(x',x) = - D^2(x,x') \ \ \ \ \ \ \text{   if $x$ and $x'$ are the two ends of a link}  
\\
0 \text{  otherwise}  
\end{cases}
\label{jdaggerj}
\end{eqnarray}

Whenever the spectral decomposition of Eq. \ref{SVDdaggerJJ} is known
 for the matrix ${\bold J}^{\dagger}{\bold J} $,
 then the $N$ singular values $\lambda_{\beta=1,..,N}>0$  
 and the $N$ left singular vectors $\vert \lambda^L_{\beta=1,..,N} \rangle $ of the matrix ${\bold J}$ are known,
while the corresponding $N$ right singular vectors $ \vert \lambda^R_{\beta=1,..,N} \rangle \! \rangle$ 
can be obtained via the application of the matrix ${\bold J} $ to the 
left singular vectors $\vert \lambda^L_{\beta=1,..,N} \rangle $
\begin{eqnarray}
{\bold J} \vert \lambda^L_{\beta} \rangle  &&  
= \lambda_{\beta} \vert \lambda^R_{\beta} \rangle  \! \rangle
\label{RfromLJ}
\end{eqnarray}

%%%%%%%%%%%%%%%%%%%%%%%%%%%%%%%%%%%%%%

\subsection{ Analog of the Helmholtz decomposition when the incidence matrix ${\bold I}$ is replaced by the current matrix ${\bold J}$} 

\label{subsec_helmForJ}

As discussed in detail in subsection \ref{subsec_discreteHelm},
the discrete Helmholtz decomposition is directly related to the SVD of the incidence matrix ${\bold I}$.
In the present subsection, it is thus useful to describe the analog of the discrete Helmholtz decomposition when the incidence matrix ${\bold I}$ is replaced by the current matrix ${\bold J}$
using its SVD decomposition of Eq. \ref{SVDJ} as follows.

An arbitrary vector $\vert u \rangle  \! \rangle $ in the space of dimension $M$ 
can be decomposed with respect to the orthonormalized basis $\vert \lambda^R_{\beta} \rangle  \! \rangle $ of Eq. \ref{orthoJR}
in terms of its $M$ coefficients 
  \begin{eqnarray}
u_{\beta} \equiv  \langle \! \langle \lambda^R_{\beta} \vert u \rangle  \! \rangle \ \ \text{for } \beta=-(C-2),...,-1,0,+1,..,N
\label{coefsvbeta}
\end{eqnarray}
Let us now analyze the splitting of these $M$ terms into two orthogonal contributions 
of dimensions $N$ and $(C-1)$ respectively
  \begin{eqnarray}
\vert u \rangle  \! \rangle && = \sum_{\beta=-(C-2)}^{N} u_{\beta}
\vert \lambda^R_{\beta} \rangle  \! \rangle
\equiv \vert u^{[\lambda_.>0]} \rangle  \! \rangle + \vert u^{[\lambda_0=0]} \rangle  \! \rangle
\nonumber \\
\vert u^{[\lambda_.>0]} \rangle  \! \rangle&& \equiv  \sum_{\beta=1}^{N} 
u_{\beta}\vert \lambda^R_{\beta} \rangle  \! \rangle  
\nonumber \\
\vert u^{[\lambda_0=0]} \rangle  \! \rangle && \equiv \sum_{\beta=-(C-2)}^{0} 
u_{\beta}\vert \lambda^R_{\beta} \rangle  \! \rangle 
\label{JHelmhotz}
\end{eqnarray}
with the following properties.

%%%%%%%%%%%%%%%%%%%%%%%%%%%%%%%%%%%%%%%%%%

\subsubsection{ The component $\vert u^{[\lambda_0=0]} \rangle  \! \rangle $ of dimension $(C-1)$ 
associated to the 
vanishing singular value $\lambda_0=0 $ }

The component $\vert u^{[\lambda_0=0]} \rangle  \! \rangle $ of dimension $(C-1)$ written in Eq. \ref{JHelmhotz} is associated to the 
vanishing singular value $\lambda_0=0 $
and is thus annihilated by the adjoint matrix ${\bold J}^{\dagger}$ whose SVD is given in Eq. \ref{SVDJ}
  \begin{eqnarray} 
0 = {\bold J}^{\dagger} \vert u^{[\lambda_0=0]} \rangle  \! \rangle 
  \label{annihJdagger}
\end{eqnarray}

%%%%%%%%%%%%%%%%%%%%%%%%%%%%%%%%%%%%%%%%%%%%%%%%%%%%

\subsubsection{ The component $\vert u^{[\lambda_.>0]} \rangle  \! \rangle $ of dimension $N$ associated to the 
strictly positive singular values $\lambda^R_{\beta=1,..,N} >0$  }

The component $\vert u^{[\lambda_.>0]} \rangle  \! \rangle $ of Eq. \ref{JHelmhotz}
associated to the 
$N$ strictly positive singular values $\lambda^R_{\beta=1,..,N} >0 $
can be rewritten using Eq. \ref{RfromLJ}
\begin{eqnarray}
\vert u^{[\lambda_.>0]} \rangle  \! \rangle&& \equiv  \sum_{\beta=1}^{N} 
u_{\beta}\vert \lambda^R_{\beta} \rangle  \! \rangle  
=   \sum_{\beta=1}^{N} 
\frac{ u_{\beta} }{\lambda_{\beta}} 
{\bold J} \vert \lambda^L_{\beta} \rangle
= {\bold J}  \left( \sum_{\beta=1}^{N} 
\frac{ u_{\beta} }{\lambda_{\beta}}  \vert \lambda^L_{\beta} \rangle \right) \equiv {\bold J}\vert k^{[\lambda_.>0]} \rangle 
\label{uGradbasisRfromL}
\end{eqnarray}
as the application of the current matrix ${\bold J}  $ of Eq. \ref{incidencegradientmatrix}
to the following ket $\vert k^{[\lambda_.>0]} \rangle $ that lives in the space of the $N$ configurations
\begin{eqnarray}
\vert k^{[\lambda_.>0]} \rangle    \equiv \sum_{\beta=1}^{N} 
\frac{ u_{\beta} }{\lambda_{\beta}}  \vert \lambda^L_{\beta} \rangle 
\label{ketku}
\end{eqnarray}

The application of the adjoint matrix ${\bold J}^{\dagger}$ on the vector $\vert u \rangle  \! \rangle $ of Eq. \ref{JHelmhotz}
only involves the application on the component $\vert u^{[\lambda_.>0]} \rangle  \! \rangle =  {\bold J}\vert k^{[\lambda_.>0]} \rangle$ of Eq. \ref{uGradbasisRfromL},
and thus reduces to the application of the matrix ${\bold J}^{\dagger} {\bold J} $ 
with the spectral decomposition of Eq. \ref{SVDdaggerJJ} on the ket $\vert k^{[\lambda_.>0]} \rangle $
  \begin{eqnarray}
{\bold J}^{\dagger} \vert u \rangle  \! \rangle = {\bold J}^{\dagger} \vert u^{[\lambda_.>0]} \rangle  \! \rangle + 0
= {\bold J}^{\dagger} {\bold J}\vert k^{[\lambda_.>0]} \rangle
= \sum_{\beta=1}^{N} 
 u_{\beta} \lambda_{\beta}  \vert \lambda^L_{\beta} \rangle 
\label{JUJJk}
\end{eqnarray}

%%%%%%%%%%%%%%%%%%%%%%%%%%%%%%%%%%%%%%%%%%%%%%%%%%%%

\subsubsection{ Conclusion on the subspace of dimension $N$ for the physical currents $\vert j_t \rangle \! \rangle =  {\bold J}  \vert p_t \rangle $  } 

\label{subsec_physical}

The above discussion shows
that the component $\vert u^{[\lambda_0=0]} \rangle  \! \rangle $  of dimension $(C-1)$ 
corresponds to the unphysical subspace orthogonal to the physical space for the currents 
$\vert j_t \rangle \! \rangle =  {\bold J}  \vert p_t \rangle $
of Eq. \ref{currentlinkop} that are obtained from 
the application of the current matrix ${\bold J}$ to a ket $\vert p_t \rangle $
of the configuration space.
This explains why the spectral decomposition of Eq. \ref{jtpropaini} only involves the 
bi-orthogonal basis of the $\langle \! \langle  i_n  \vert $ and the $\vert j_n \rangle \! \rangle $ for $n=0,1,..,(N-1)$,
even if the partner ${\hat {\bold H}} $ that governs the dynamics of the current is a matrix of size $M \times M$.

The main conclusion is thus that the projector ${\bold {\cal P}}^{PhysicalSpaceCurrents} $ 
onto the subspace of dimension $N$ for the physical currents
can be written either 
with the bi-orthogonal basis of the $\langle \! \langle  i_n  \vert $ and the $\vert j_n \rangle \! \rangle $ for $n=0,1,..,(N-1)$
or with the orthonormalized basis $\vert \lambda^R_{\beta=1,..,N} \rangle  \! \rangle $ of the right singular vectors of 
the current matrix ${\bold J}$ associated to the 
strictly positive singular values $\lambda^R_{\beta=1,..,N} >0$
 \begin{eqnarray}
{\bold {\cal P}}^{PhysicalSpaceCurrents}   = \sum_{n=0}^{N-1}    \vert j_n \rangle \! \rangle 
\langle \! \langle  i_n  \vert 
 = \sum_{\beta=1}^{N  }  \vert \lambda^R_{\beta} \rangle \! \rangle \langle \! \langle \lambda^R_{\beta} \vert
\label{ProjectorPhysicalCurrents}
\end{eqnarray}
while the orthogonal projector involving the other $(C-1)$ right singular vectors 
$\vert \lambda^R_{\beta=-(C-2),..,0} \rangle \! \rangle $ 
associated to the vanishing singular value $\lambda_0=0$
 \begin{eqnarray}
{\bold {\cal P}}^{UnhysicalSpaceCurrents}    
= \sum_{\beta=-(C-2)}^{0}   \vert \lambda^R_{\beta} \rangle \! \rangle
 \langle \! \langle \lambda^R_{\beta} \vert
\label{ProjectorUNPhysicalCurrents}
\end{eqnarray}
corresponds to the unphysical subspace for the currents.

%%%%%%%%%%%%%%%%%%%%%%%%%%%%%%%%%%%%%%%%%

\subsubsection{ Application to the definition of the pseudo-inverse ${\bold J}^{pseudo[-1]}  $ of the current matrix ${\bold J}  $}

The linear system involving the rectangular current matrix ${\bold J}$ with the SVD of Eq. \ref{SVDJ}
\begin{eqnarray}
\vert u \rangle \! \rangle = {\bold J} \vert k \rangle \equiv  \sum_{\beta=1}^{N  } \lambda_{\beta} 
\vert \lambda^R_{\beta} \rangle  \! \rangle \langle \lambda^L_{\beta} \vert k \rangle
\label{linearsystemJ}
\end{eqnarray}
for the unknown ket $\vert k \rangle $ of dimension $N$ 
when the vector $\vert u \rangle \! \rangle$ of dimension $M$ is given,
can be analyzed via the following adaptation of
the discussion of subsection \ref{subsec_pseudoInversionI} :

(i) The $(C-1)$ coefficients $ u_{\beta=0,-1,..,-(C-2)} $ should vanish
\begin{eqnarray}
u_{\beta} \equiv  \langle \! \langle \lambda^R_{\beta} \vert u \rangle  \! \rangle =0  \ \ \text{ for } \beta=0,-1,..,-(C-2)
\label{linearsystemJSDVzerocoefs}
\end{eqnarray}

(ii) The solution for the ket $\vert  k \rangle $
  \begin{eqnarray}
\vert  k \rangle  = {\bold J}^{pseudo[-1]} \vert u \rangle  \! \rangle
\label{linearsystemJSDVsolution}
\end{eqnarray}
corresponds to the application to the vector $\vert u \rangle  \! \rangle $
of the pseudo-inverse ${\bold J}^{pseudo[-1]} $ of the current matrix ${\bold J} $ with the SVD of Eq. \ref{SVDJ}
\begin{eqnarray}
{\bold J}^{pseudo[-1]} \equiv  \sum_{\beta=1}^{N  } \frac{1}{\lambda_{\beta} }
\vert \lambda^L_{\beta} \rangle \langle \! \langle \lambda^R_{\beta} \vert
\label{SVDJpseudo}
\end{eqnarray}

%%%%%%%%%%%%%%%%%%%%%%%%%%%%%%%%%%%%%%%%

\subsection{ Solving the linear system $\vert j_* \rangle \! \rangle  = {\bold J} \vert p_* \rangle $
 to obtain the steady state $p_*$ and the current $j_*$}

\label{subsec_pseudoinverseJstar}

Let us consider the application of the previous subsection
to the linear system of Eq. \ref{jjPbasislink}
\begin{eqnarray}
\vert j_* \rangle \! \rangle = {\bold J} \vert p_* \rangle \equiv  \sum_{\beta=1}^{N  } \lambda_{\beta} 
\vert \lambda^R_{\beta} \rangle  \! \rangle \langle \lambda^L_{\beta} \vert p_* \rangle
\label{jstarpstar}
\end{eqnarray}
for the unknown steady state $\vert p_* \rangle $ in the space of the $N$ configurations,
once the steady current $\vert j_* \rangle \! \rangle $ has been parametrized by its $C$ coefficients in Eq. \ref{jsteadycyclesGamma} :

(i) Eq. \ref{linearsystemJSDVzerocoefs}
gives $(C-1)$ equations for the $C$ coefficients parametrizing the steady current $\vert j_* \rangle \! \rangle $
\begin{eqnarray}
 \langle \! \langle \lambda^R_{\beta} \vert j_* \rangle \! \rangle  = 0
\ \ \text{ for } \beta=0,-1,..,-(C-2)
\label{jstarprjSVDZero}
\end{eqnarray}

(ii) Eq. \ref{linearsystemJSDVsolution}
gives the steady state of dimension $N$
\begin{eqnarray}
\vert p_* \rangle = {\bold J}^{pseudo[-1]} \vert j_* \rangle \! \rangle  
= \sum_{\beta=1}^{N  } \frac{1}{\lambda_{\beta} }
\vert \lambda^L_{\beta} \rangle \langle \! \langle \lambda^R_{\beta} \vert j_* \rangle \! \rangle
\label{jpstarjstar}
\end{eqnarray}
in terms of the projections of the steady current $\vert j_* \rangle \! \rangle $ 
on the $N$ singular eigenvectors $\langle \! \langle \lambda^R_{\beta=1,..,N} \vert $.
Then the normalization of the steady state $p_*(.)$ of Eq. \ref{jpstarjstar}
determines the last remaining unknown coefficient for the steady current.

%%%%%%%%%%%%%%%%%%%%%%%%%%%%%%%%%%

\subsection{ Consequences for the excited right eigenstates $ \vert r_{n=1,..,N-1} \rangle $
and their associated currents $ \vert j_n \rangle \! \rangle  = {\bold J} \vert r_n \rangle $
 }

Plugging the SVD of ${\bold J} $ of Eq. \ref{SVDJ} into Eq. \ref{jn}
\begin{eqnarray}
 \vert j_n \rangle \! \rangle && = {\bold J} \vert r_n \rangle
 =
  \sum_{\beta=1}^{N  } \lambda_{\beta} 
\vert \lambda^R_{\beta} \rangle  \! \rangle \langle \lambda^L_{\beta} \vert   r_n \rangle
\label{spectralrsvdj}
\end{eqnarray}
yields that the excited right eigenstate $ \vert r_{n=1,..,N-1} \rangle $
can be also computed from the current $\vert j_n \rangle \! \rangle $
via the pseudo-inverse ${\bold J}^{pseudo[-1]}  $  as in Eq. \ref{jpstarjstar}
\begin{eqnarray}
\vert r_n \rangle = {\bold J}^{pseudo[-1]} \vert j_n \rangle \! \rangle  
= \sum_{\beta=1}^{N  } \frac{1}{\lambda_{\beta} }
\vert \lambda^L_{\beta} \rangle \langle \! \langle \lambda^R_{\beta} \vert j_n \rangle \! \rangle
\label{rnpseudojn}
\end{eqnarray}

%%%%%%%%%%%%%%%%%%%%%%%%%%%%%%%

\subsection{ Consequences for the left eigenstates $   \langle l_{n=0,..,N-1} \vert $
and their associated $\langle \! \langle i_n \vert  $  satisfying $   \langle l_n \vert  =  \langle \! \langle i_n \vert {\bold J} $
 }

Plugging the SVD of ${\bold J} $ of Eq. \ref{SVDJ} into Eq. \ref{spectrallin} yields
\begin{eqnarray}
  \langle l_n \vert && =  \langle \! \langle i_n \vert {\bold J} 
 = 
  \sum_{\beta=1}^{N  } \lambda_{\beta} 
\langle \! \langle i_n \vert \lambda^R_{\beta} \rangle  \! \rangle \langle \lambda^L_{\beta} \vert
\label{spectrallsvdj}
\end{eqnarray}
Since $\langle \! \langle i_n \vert $ belongs to the subspace of physical currents of Eq. \ref{ProjectorPhysicalCurrents},
Eq. \ref{spectrallsvdj} can be inversed to compute $\langle \! \langle i_n \vert $ in terms of $ \langle l_n \vert $ via
 \begin{eqnarray}
\langle \! \langle  i_n  \vert 
&& = \sum_{\beta=1}^{N  } \langle \! \langle  i_n  \vert  \vert \lambda^R_{\beta} \rangle \! \rangle 
\langle \! \langle \lambda^R_{\beta} \vert
= \sum_{\beta=1}^{N  } \frac{ \langle  l_n  \vert   \lambda^L_{\beta} \rangle }{\lambda_{\beta} } 
\langle \! \langle \lambda^R_{\beta} \vert = \langle  l_n  \vert  {\bold J}^{pseudo[-1]}
\label{infromln}
\end{eqnarray}
that involves the pseudo-inverse $  {\bold J}^{pseudo[-1]}$.

For $n=0$ with the left eigenvector $l_0(x)=1$, Eq. \ref{infromln}
yields that the bra $ \langle \! \langle  i_0  \vert$ can be evaluated from the pseudo-inverse $  {\bold J}^{pseudo[-1]}$ via
 \begin{eqnarray}
\langle \! \langle  i_0  \vert && = \langle  l_0  \vert  {\bold J}^{pseudo[-1]}
= \sum_x  \langle  l_0  \vert x \rangle \langle x \vert  {\bold J}^{pseudo[-1]}
\nonumber \\
&&  = \sum_x   \langle x \vert  {\bold J}^{pseudo[-1]}
= \sum_{\beta=1}^{N  } \frac{ \langle  x  \vert   \lambda^L_{\beta} \rangle }{\lambda_{\beta} } 
\langle \! \langle \lambda^R_{\beta} \vert 
\label{i0froml0}
\end{eqnarray}

%%%%%%%%%%%%%%%%%%%%%%%%%%%%%%%%%%%%

 \subsection{ Discussion   }

In summary, the Singular Value Decomposition of the current matrix 
${\bold J} $ of size $M \times N$ is very useful to characterize 
the subspace of dimension $N$ for the physical currents 
via the projector of Eq. \ref{ProjectorPhysicalCurrents},
while the definition of the pseudo-inverse ${\bold J}^{pseudo[-1]} $
clarifies the relations between the biorthogonal basis 
of right and left eigenvectors of the Hamiltonian ${\bold H}={\bold I}^{\dagger}{\bold J}$ and of its supersymmetric partner ${\hat{\bold H}}= {\bold J}{\bold I}^{\dagger}$.

The general analysis of spectral properties of non-equilibrium Markov jump processes 
is now finished for the present paper : the application to a simple translation-invariant model
can be found in Appendix \ref{app_example}, 
while the next section of the main text is devoted to the comparison 
with the spectral properties of non-equilibrium diffusion processes in dimension $d=3$.

%%%%%%%%%%%%%%%%%%%%%%%%%%%%%%%%%%%%%%%%

\section{  Comparison with spectral properties of Fokker-Planck generators }

\label{app_diffusion}

For diffusion processes in dimension $d$, 
the essential ideas are the same as in the previous sections concerning Markov jump processes,
but there are some important technical differences :

(i) the number $N$ of configurations, the number $M$ of links, 
and the number $C$ of independent cycles become infinite, so there is the complication that
 the different spaces and subspaces discussed previously become all infinite-dimensional.
 
(ii) the incidence matrix ${\bold I}$ and the current matrix ${\bold J}$ of size $M \times N$ become first-order differential operators acting on scalar functions to produce vector fields,
so there is the simplification that the vector calculus operators like the gradient and the divergence, 
as well as the Helmholtz decomposition of vectors, are more familiar in continuous space than in discrete space.

For concreteness and to simplify the notations, we will focus on the spatial dimension $d=3$
in order to use the standard 3D curl operator that is simpler 
than its generalization in higher dimensions $d>3$.

%%%%%%%%%%%%%%%%%%%%%%%%%%%%%%%%%%%%%%%%%%

\subsection{Fokker-Planck dynamics in terms of the probability density  $p_t( \vec x ) $ and the 3D current $\vec j_t( \vec x ) $  }

The Fokker-Planck equation for the probability density  $p_t( \vec x ) $ to be around the position $\vec x$ at time $t$
can be written as the continuity equation 
 \begin{eqnarray}
 \partial_t p_t( \vec x )  = - {\bold {div}} \  \vec j_t( \vec x )  
\label{fokkerplanck}
\end{eqnarray}
where the 3D current $\vec j_t( \vec x ) $ involves the force $\vec F(\vec x) $ and the diffusion coefficient $ D(\vec x) $
\begin{eqnarray}
\vec j_t( \vec x ) = \vec F(\vec x)  p_t(x) - D(\vec x) \vec{{\bold {grad}}} \ p_t( \vec x ) 
\label{jfokkerplanck}
\end{eqnarray}

%%%%%%%%%%%%%%%%%%%%%%%%%%%%%%%%%%%%%%%%%%

\subsection{  Rephrasing with the first-order differential operators ${\bold J}  $ and ${\bold I}  $ }

The analog of the current matrix ${\bold J} $ of size $M \times N$ of Eqs \ref{currentOpconfig} and \ref{currentlinkop}
is the first-order differential operator 
\begin{eqnarray}
 {\bold J} \equiv \vec F(\vec x)   - D(\vec x) \vec{{\bold {grad}}}
=  \begin{pmatrix} 
   F_1(\vec x)   - D(\vec x) \frac{\partial}{\partial x_1} 
 \\  
  F_2(\vec x)   - D(\vec x) \frac{\partial}{\partial x_2} 
 \\ 
  F_3(\vec x)   - D(\vec x) \frac{\partial}{\partial x_3} 
  \end{pmatrix}
\label{Jopfokkerplanck}
\end{eqnarray}
 that acts on the scalar density $p_t(x) $ 
to produce the 3D current $\vec j_t( \vec x ) $ of Eq. \ref{jfokkerplanck}
\begin{eqnarray}
\vec j_t( \vec x ) =  {\bold J}  p_t(x)
\label{jfokkerplanckop}
\end{eqnarray}

The analog of the incidence matrix $ {\bold I} $ of size $M \times N$ of Eq. \ref{incidence} \ref{incidencegradientmatrix}
is the opposite of the gradient operator,
that can be recovered from the current operator of Eq. \ref{Jopfokkerplanck}
for the simplest case where the force vanishes $\vec F(\vec x) \to 0 $ 
and where the diffusion coefficient reduces to unity $D(\vec x) \to 1 $
\begin{eqnarray}
 {\bold I} \equiv    -  \vec{{\bold {grad}}}
= -  \begin{pmatrix} 
       \frac{\partial}{\partial x_1} 
 \\  
   \frac{\partial}{\partial x_2} 
 \\ 
   \frac{\partial}{\partial x_3} 
  \end{pmatrix}
\label{Iopfokkerplanck}
\end{eqnarray}
The adjoint operator ${\bold I}^{\dagger} $ is the divergence operator
that acts on 3D vectors to produce a scalar
\begin{eqnarray}
 {\bold I}^{\dagger} \equiv      {\bold {div}}
=   \begin{pmatrix} 
       \frac{\partial}{\partial x_1} 
 \  
   \frac{\partial}{\partial x_2} 
 \ 
   \frac{\partial}{\partial x_3} 
  \end{pmatrix}
\label{Idaggerfokkerplanck}
\end{eqnarray}
which is the analog of the discrete divergence matrix
of size $N \times M$ of Eqs \ref{divergencefromincidence} and \ref{divergencefromincidencematrix}.

The Fokker-Planck Eq. \ref{fokkerplanck} can be rewritten as the euclidean Schr\"odinger equation
 \begin{eqnarray}
- \partial_t p_t( \vec x )  =  {\bold H}   p_t( \vec x )  
\label{fokkerplanckschro}
\end{eqnarray}
where the second-order differential non-hermitian Hamiltonian ${\bold H} \ne {\bold H}^{\dagger} $ 
is factorized in terms of the two first-order differential operators ${\bold I}^{\dagger}  $ and $ {\bold J}  $
\begin{eqnarray}
 {\bold H}  =  {\bold I}^{\dagger}  {\bold J} 
 =  \begin{pmatrix} 
       \frac{\partial}{\partial x_1} 
 \  
   \frac{\partial}{\partial x_2} 
 \ 
   \frac{\partial}{\partial x_3} 
  \end{pmatrix}
 \begin{pmatrix} 
   F_1(\vec x)   - D(\vec x) \frac{\partial}{\partial x_1} 
 \\  
  F_2(\vec x)   - D(\vec x) \frac{\partial}{\partial x_2} 
 \\ 
  F_3(\vec x)   - D(\vec x) \frac{\partial}{\partial x_3} 
  \end{pmatrix}
  = \sum_{\mu=1}^3 \frac{\partial}{\partial x_{\mu} } \left( F_{\mu}(\vec x)   - D(\vec x) \frac{\partial}{\partial x_{\mu}}  \right)
\label{Hfactordiff}
\end{eqnarray}

%%%%%%%%%%%%%%%%%%%%%%%%%%%%%%%%%%%%%%%%%%%

\subsection{  Electromagnetic quantum interpretation of the non-hermitian Hamiltonian ${\bold H}  $ }

The electromagnetic quantum interpretation of the non-hermitian Hamiltonian ${\bold H}  $
of Eq. \ref{Hfactordiff}
is based on the rewriting 
\begin{eqnarray}
 {\bold H}  
&&  = - \sum_{\mu=1}^3 \left( \frac{\partial}{\partial x_{\mu} } - A_{\mu}(\vec x) \right)
D(\vec x)   \left(  \frac{\partial}{\partial x_{\mu}} - A_{\mu}(\vec x) \right)
  + V(\vec x)
  \nonumber \\
  && =  \sum_{\mu=1}^3 \left( -i \frac{\partial}{\partial x_{\mu} } +i A_{\mu}(\vec x) \right)
D(\vec x)   \left( -i \frac{\partial}{\partial x_{\mu}} +i A_{\mu}(\vec x) \right)
  + V(\vec x)
   \nonumber \\
  && =   \left( -i \vec \nabla +i \vec A(\vec x) \right) D(\vec x)  \left( -i \vec \nabla +i \vec A(\vec x) \right)
  + V(\vec x)
\label{Helectromagn}
\end{eqnarray}
that involves a purely imaginary vector potential $[ -i \vec A(\vec x) ] $ of real amplitude
\begin{eqnarray}
\vec A(\vec x) \equiv \frac{ \vec F(\vec x) }{ 2 D(\vec x) } 
 \label{Avectorpot}
\end{eqnarray}
which is the analog of the antisymmetric function $A(.,.)$ of Eq. \ref{Avecpot}
that appears in the off-diagonal terms ${\bold H} (x,x') $ of Eq. \ref{Hwrates},
while the scalar potential
\begin{eqnarray}
V(\vec x) && \equiv \sum_{\mu=1}^3  \left( D(\vec x) A^2_{\mu}(\vec x) +   \frac{\partial [D(\vec x)A_{\mu}(\vec x) ] }{\partial x_{\mu}}  \right)
\nonumber \\
&& = \sum_{\mu=1}^3  \left( \frac{ F^2_{\mu}(\vec x) }{ 4 D(\vec x) } + \frac{1}{2}  \frac{\partial F_{\mu}(\vec x)  }{\partial x_{\mu}}  \right)
 \label{Vscalarpot}
\end{eqnarray}
is the analog of the on-site potential ${\bold H} (x,x) $ of Eq. \ref{Hwdiag}.

The magnetic field $\vec B  ( \vec x ) $ associated to the vector potential $\vec A ( \vec x ) $ of Eq. \ref{Avectorpot} 
\begin{eqnarray}
\vec B  ( \vec x ) \equiv {\bold {curl} } \ \vec A ( \vec x ) = \vec \nabla \times \vec A ( \vec x )
=  \begin{pmatrix} 
       \frac{\partial A_3}{\partial x_2} - \frac{\partial A_2}{\partial x_3} 
 \\
    \frac{\partial A_1}{\partial x_3} - \frac{\partial A_3}{\partial x_1} 
 \\ 
   \frac{\partial A_2}{\partial x_2} - \frac{\partial A_1}{\partial x_2}  
  \end{pmatrix}
\label{Bmagnetic}
\end{eqnarray}
is useful to rewrite the circulation of the vector potential $\vec A ( \vec x ) $ around any closed curve $\gamma$
\begin{eqnarray} 
\Gamma^{[\gamma]}[\vec A(.)]  \equiv \oint_{\gamma} d \vec l . \vec A ( \vec x ) = \int d^2 \vec S . \vec B  ( \vec x )
  \label{stokes}
\end{eqnarray}
as the flux of the magnetic field $\vec B  ( \vec x ) $ through the surface enclosed by the closed curve $\gamma$.
This magnetic field $\vec B  ( \vec x ) $ determines the equilibrium or non-equilibrium nature of the steady state as  we now recall (see \cite{us_gyrator} for more detailed discussions).

 %%%%%%%%%%%%%%%%%%%%%%%%%%%%%%%%%%%%%%%%%%
 
 \subsection{ Reminder on the properties of the steady current $\vec j_* (.) $}

 \subsubsection{ Equilibrium steady state $p_*^{eq}(.) $ with vanishing steady currents $\vec j^{eq}_*(\vec x)=\vec 0$}

At equilibrium, the steady current $\vec j^{eq}_*(\vec x) $ associated to the steady density $p_*^{eq}(.)  $
 vanishes everywhere
\begin{eqnarray}
\vec 0 = \vec j^{eq}_*( \vec x ) = \vec F(\vec x)  p_*^{eq}(x) - D(\vec x) \vec \nabla p_*^{eq}( \vec x ) 
\label{jeqfokkerplanck}
\end{eqnarray}
This is possible only if the vector potential $\vec A \equiv \frac{ \vec F(\vec x)  }{ 2 D(\vec x) } $ introduced in Eq. \ref{Avectorpot}
can be written as the gradient 
\begin{eqnarray}
\vec A ( \vec x )= \frac{1}{2} \vec \nabla \ln p_*^{eq}( \vec x ) 
\label{Agradfokkerplanck}
\end{eqnarray}
This condition is the continuous analog of Eq. \ref{Agradient}
and means that the circulation of the vector potential $\vec A ( \vec x ) $ of Eq. \ref{stokes}
vanishes along any closed curve $\gamma$ as in Eq. \ref{KolmogorovcirculationA}
\begin{eqnarray} 
\Gamma^{[\gamma]}[\vec A(.) ]  \equiv \oint_{\gamma} d \vec l . \vec A ( \vec x ) =0
  \label{KolmogorovcirculationAdiff}
\end{eqnarray}
i.e. that the magnetic field $ \vec B ( \vec x ) \equiv \vec \nabla \times \vec A ( \vec x )$ 
introduced in Eq. \ref{Bmagnetic} vanishes everywhere
\begin{eqnarray}
\vec B ( \vec x ) \equiv \vec \nabla \times \vec A ( \vec x )= \vec 0
\label{Bvanish}
\end{eqnarray}

%%%%%%%%%%%%%%%%%%%%%%%%%%%%%%%%%%%%%%%%%%%%%%%%%%%%%

\subsubsection{ Non-equilibrium steady state $p_*(.) $ with nonvanishing steady currents $\vec j_*(\vec x)=\vec 0$}

When the magnetic field $ \vec B ( \vec x ) \equiv \vec \nabla \times \vec A ( \vec x )$ 
associated to the vector potential $\vec A ( \vec x ) \equiv \frac{ \vec F(\vec x)  }{ 2 D(\vec x) } $  of Eq. \ref{Avectorpot}
 does not vanish
 \begin{eqnarray}
  \vec B ( \vec x ) \equiv \vec \nabla \times \vec A ( \vec x ) \ne \vec 0
\label{Bnonvanish}
\end{eqnarray}
 then the steady current 
\begin{eqnarray}
 \vec j_*( \vec x ) \equiv  \vec F(\vec x)  p_*(x) - D(\vec x) \vec \nabla  p_*( \vec x ) 
 =
2 D(\vec x) p_*(x)\left[   \vec A(\vec x)  - \frac{1}{2} \vec \nabla \ln p_*( \vec x ) \right] \ne \vec 0 
\label{jnoneqfokkerplanck}
\end{eqnarray}
cannot vanish,
but should be divergenceless  
\begin{eqnarray}
\text{div} \vec j_*( \vec x ) =0
\label{divjnoneqfokkerplanck}
\end{eqnarray}
So the steady current $ \vec j_*( \vec x ) $ can be rewritten as the curl of another divergenceless vector $\vec \omega_* ( \vec x ) $
\begin{eqnarray}
 \vec j_*( \vec x )  && = \vec \nabla \times \vec \omega_* ( \vec x )
 \nonumber \\
\bold{div} \  \vec \omega_* ( \vec x ) && \equiv \vec \nabla . \vec \omega_* ( \vec x )  =0
\label{jnoneqcurl}
\end{eqnarray}
which is the analog of the decomposition into cycles-currents of Eq. \ref{jsteadycyclesGamma}.
The curl of the steady current
\begin{eqnarray}
\vec \nabla \times \vec j_*( \vec x )   = \vec \nabla \times \left( \vec \nabla \times \vec \omega_* ( \vec x ) \right)
= \vec \nabla \left( \vec \nabla . \vec \omega_* ( \vec x ) \right) - \Delta \vec \omega_* ( \vec x )
= - \Delta \vec \omega_* ( \vec x )
\label{curljnoneqcurl}
\end{eqnarray}
corresponds to the opposite Laplacian of the vector $\vec \omega_* ( \vec x ) $.

%%%%%%%%%%%%%%%%%%%%%%%%%%%%

\subsection{ Spectral decomposition of the Hamiltonian ${\bold H} $ in the bi-orthogonal basis of 
right and left eigenvectors }

In order to simplify the discussion and the notations, 
we will consider that the Fokker-Planck dynamics takes place in a finite domain $\vec x \in {\cal V}$ 
with reflecting boundary conditions, 
so that
the spectral decomposition of the non-hermitian Hamiltonian ${\bold H} \ne {\bold H}^{\dagger}$
of Eq. \ref{Hfactordiff} does not involve a continuum of eigenvalues,
but only an infinite series of discrete eigenvalues $E_{n=0,1,2,..}$ 
(instead of the finite number $N$ of eigenvalues of Eq. \ref{spectral} for Markov jump processes
in a space of $N$ configurations) 
\begin{eqnarray}
{\bold H} =  \sum_{n=0}^{+\infty} E_n \vert r_n \rangle \langle l_n \vert
\label{spectraldiff}
\end{eqnarray}
The corresponding right eigenvectors $ r_n (\vec x)= \langle  \vec x \vert r_n \rangle $, 
and the corresponding left eigenvectors $ l_n (\vec x)= \langle \vec x \vert l_n \rangle  $, 
that are equivalently the right eigenvectors of the adjoint operator $ {\bold H}^{\dagger}$
associated to the complex-conjugate eigenvalue $E_n^*$,
satisfy the eigenvalue equations
\begin{eqnarray}
  E_n  r_n (\vec x)&& =  {\bold H}  r_n (\vec x) = 
  \sum_{\mu=1}^3 \frac{\partial}{\partial x_{\mu} } \left( F_{\mu}(\vec x)  r_n (\vec x)  - D(\vec x) \frac{\partial  r_n (\vec x)}{\partial x_{\mu}}  \right)
  \nonumber \\
 E_n^*  l_n  (\vec x) && = {\bold H}^{\dagger}   l_n  (\vec x) =
-  \sum_{\mu=1}^3 
  \left( F_{\mu}(\vec x)   +  \frac{\partial}{\partial x_{\mu}}  D(\vec x) \right)
  \frac{\partial  l_n  (\vec x)}{\partial x_{\mu} }  
\label{spectralrldiff}
\end{eqnarray}
and form a bi-orthogonal basis where the orthonormalization and closure relations of Eq. \ref{orthorl} become
\begin{eqnarray}
\delta_{n,n'} && = \langle l_n \vert  r_{n'} \rangle = \int_{\cal V} d^3 \vec x 
\langle l_n \vert \vec x \rangle \langle \vec x \vert r_{n'} \rangle 
=  \int_{\cal V} d^3 \vec x  l_n^* (\vec x)  r_n (\vec x')
\nonumber \\
\delta^{(3)}( \vec x - \vec x ')  && = \langle  \vec x \vert \vec x ' \rangle
= \sum_{n=0}^{+\infty} \langle  \vec x \vert r_n \rangle \langle l_n \vert  \vec x ' \rangle
= \sum_{n=0}^{+\infty}  r_n (\vec x)  l_n^* (\vec x')
\label{orthorldiff}
\end{eqnarray}
As in Eq. \ref{rlzero}, 
the vanishing eigenvalue $E_0=0$ is associated to the positive left eigenvector unity
and to the positive right eigenvector given by the steady density $p_*(.)$
\begin{eqnarray}
  E_0 && =0
  \nonumber \\
 l_0 ( \vec x)&& =1
  \nonumber \\
 r_0 (\vec x)&& =p_*( \vec x)
\label{rlzerodiff}
\end{eqnarray}
while the other eigenvalues $E_{n=1,..,+\infty}$ with strictly positive real parts ${\text Re}(E_n)>0$
govern the relaxation towards the steady density $p_*(x)$ as in Eqs \ref{propagator} and \ref{observable}.

At the level of the eigenvalues Eq. \ref{spectralrldiff},
the factorization $\bold{H}={\bold I}^{\dagger}{\bold J}$  of Eq. \ref{Hfactordiff}
 corresponds for the right eigenvectors $ r_n (\vec x) $
to the splitting into the pair of two first-order differential equations
\begin{eqnarray}
  E_n r_n (\vec x) && =   {\bold I}^{\dagger}   \vec  j_n (\vec x)  =  \vec \nabla .  \vec  j_n (\vec x) 
    \nonumber \\
 \vec  j_n (\vec x)  && = {\bold J} r_n (\vec x) = \left[ \vec F(\vec x)   - D(\vec x) \vec \nabla \right] r_n (\vec x)
\label{spectralrrdiff}
\end{eqnarray}
that is the analog of Eqs \ref{jn} \ref{spectralrnjn},
while the corresponding splitting for the left eigenvectors $ l_n (\vec x)  $
\begin{eqnarray}
 l_n  (\vec x) && = {\bold J}^{\dagger}  \vec  i_n (\vec x)=
   \vec F(\vec x) . \vec  i_n (\vec x)  + \vec \nabla . \left( D(\vec x) . \vec  i_n (\vec x) \right)     
\nonumber \\
 E_n^*  \vec  i_n (\vec x)  && = {\bold I} l_n (\vec x) =  -  \vec \nabla l_n (\vec x) 
\label{spectralldiff}
\end{eqnarray}
is the analog of Eq \ref{spectrallin}.

%%%%%%%%%%%%%%%%%%%%%%%%%%%%

\subsection{ Dynamics of the current $ \vec j_t(\vec x)$ governed by the supersymmetric partner $\hat {\bold H}  =  {\bold J}  {\bold I}^{\dagger}  $}

As in Eq. \ref{mastereqschroj}, 
the current $ \vec j_t(\vec x)$ satisfies the closed dynamics
\begin{eqnarray}
- \partial_t \vec j_t(\vec x) && = {\bold J} \bigg( - \partial_t \vert p_t \rangle \bigg) 
= {\bold J} {\bold I}^{\dagger} \vec j_t(\vec x) 
\equiv {\hat {\bold H} } \vec j_t(\vec x) 
\label{diffschroj}
\end{eqnarray}
governed by the supersymmetric partner $\hat {\bold H}  =  {\bold J}  {\bold I}^{\dagger}  $ of the Hamiltonian $ {\bold H}  =  {\bold I}^{\dagger}  {\bold J}  $ of Eq. \ref{Hfactordiff}
\begin{eqnarray}
\hat {\bold H}  =  {\bold J}  {\bold I}^{\dagger} 
 =  \begin{pmatrix} 
   F_1(\vec x)   - D(\vec x) \frac{\partial}{\partial x_1} 
 \\  
  F_2(\vec x)   - D(\vec x) \frac{\partial}{\partial x_2} 
 \\ 
  F_3(\vec x)   - D(\vec x) \frac{\partial}{\partial x_3} 
  \end{pmatrix}
  \begin{pmatrix} 
       \frac{\partial}{\partial x_1} 
 \  
   \frac{\partial}{\partial x_2} 
 \ 
   \frac{\partial}{\partial x_3} 
  \end{pmatrix}
  = \bigg[ \left( F_{\mu}(\vec x)   - D(\vec x) \frac{\partial}{\partial x_{\mu}}  \right) \frac{\partial}{\partial x_{\nu} }\bigg]_{\mu=1,2,3;\nu=1,2,3}
\label{partnerHfokkerplanck}
\end{eqnarray}
that corresponds to a $3 \times 3$ matrix of differential operators 
acting on the 3 components of the current.

As discussed around between Eqs \ref{jtpropa} and \ref{jtpropaini}, 
the dynamics of the current $\vec j_t(\vec x) $
involves the same non-vanishing eigenvalues $E_{n=1,..,+\infty}$ as the hamiltonian ${\bold H} $,
where the right and left eigenvectors of the partner $\hat {\bold H} $
are given by the $\vec  j_n (\vec x)$ of Eq. \ref{spectralrrdiff}
and $\vec  i_n (\vec x) $ of Eq. \ref{spectralldiff}
\begin{eqnarray}
E_n \vec  j_n (\vec x)  && = {\bold J} {\bold I}^{\dagger}   \vec  j_n (\vec x) =\hat {\bold H} \vec  j_n (\vec x)
\nonumber \\
 E_n^*  \vec  i_n (\vec x)  && = {\bold I}  {\bold J}^{\dagger}  \vec  i_n (\vec x) = \hat {\bold H}^{\dagger} \vec  i_n (\vec x)
\label{eigenpartnerdiff}
\end{eqnarray}

For $n=0$, the steady current $\vec  j_{n=0} (\vec x) = \vec  j_* (\vec x) $ is annihilated by 
the divergence operator $ {\bold I}^{\dagger}=\bold {div} $ and thus by the partner $\hat {\bold H} $
\begin{eqnarray}
  {\bold I}^{\dagger}   \vec  j_* (\vec x) =0 = \hat {\bold H} \vec  j_* (\vec x)
\label{eigenpartnerdiffjstar}
\end{eqnarray}
while $\vec  i_0 (\vec x) $ satisfies Eq. \ref{spectralldiff} with $l_0  (\vec x)=1$ annihilated by 
$ {\bold I}=- \bold {grad} $
\begin{eqnarray}
1 && = l_0  (\vec x)  = {\bold J}^{\dagger}  \vec  i_0 (\vec x)=
   \vec F(\vec x) . \vec  i_0 (\vec x)  + \vec \nabla . \left( D(\vec x) . \vec  i_0 (\vec x) \right)     
\nonumber \\
 0  && = {\bold I} l_0  (\vec x) = {\bold I}  {\bold J}^{\dagger}  \vec  i_0 (\vec x) =\hat {\bold H}^{\dagger} \vec  i_0 (\vec x)
\label{spectralldiffzero}
\end{eqnarray}

%%%%%%%%%%%%%%%%%%%%%%%%%%%%%%%%%%%%%%%%%%%%%%%%%%%%%%

\subsection{ Singular Value Decompositions for the differential operator ${\bold I} =-  \vec{{\bold {grad}}}$ and its adjoint $ {\bold I}^{\dagger} \equiv      {\bold {div}} $ }

The Singular Value Decomposition for the operator ${\bold I} =-  \vec{{\bold {grad}}}$ and its adjoint $ {\bold I}^{\dagger} \equiv      {\bold {div}} $
 involves an infinite series of strictly positive singular values $ I_{\alpha}>0 $
 (instead of the $(N-1)$ values in Eq. \ref{SVDI} concerning Markov jump processes in a space of $N$ configurations)
\begin{eqnarray}
{\bold I} && \equiv -  \vec{{\bold {grad}}} =  \sum_{\alpha=1}^{+\infty  } I_{\alpha} \vert I^R_{\alpha} \rangle  \! \rangle \langle I^L_{\alpha} \vert
\nonumber \\
{\bold I}^{\dagger} &&\equiv      {\bold {div}}  =  \sum_{\alpha=1}^{+\infty  } I_{\alpha} \vert I^L_{\alpha} \rangle \langle \! \langle I^R_{\alpha} \vert
\label{SVDIDiff}
\end{eqnarray}

As in Eq. \ref{Izero}, the vanishing singular value
\begin{eqnarray}
 I_{(\alpha=0)} =0
\label{Izerodiff}
\end{eqnarray}
is associated to the uniform normalized eigenvector $ \langle x \vert I^L_{(\alpha=0)} \rangle$ over the bounded volume 
$\vec x \in {\cal V}$ where the Fokker-Planck dynamics takes place.
 The operator ${\bold I}^{\dagger}{\bold I} $ corresponds to the opposite of the Laplacian
\begin{eqnarray}
{\bold I}^{\dagger}{\bold I} && = {\bold {div}}(- {\bold {grad}} ) 
= -  \begin{pmatrix} 
       \frac{\partial}{\partial x_1} 
 \  
   \frac{\partial}{\partial x_2} 
 \ 
   \frac{\partial}{\partial x_3} 
  \end{pmatrix}
 \begin{pmatrix} 
    \frac{\partial}{\partial x_1} 
 \\  
   \frac{\partial}{\partial x_2} 
 \\ 
   \frac{\partial}{\partial x_3} 
  \end{pmatrix}
  = - \sum_{\mu=1}^3 \frac{\partial^2}{\partial x_{\mu}^2 }
=- {\bold \Delta}
\label{Laplaciandiff}
\end{eqnarray}
while its evaluation from the SVD decompositions of Eq. \ref{SVDIDiff}
\begin{eqnarray}
- {\bold \Delta} ={\bold I}^{\dagger} {\bold I} && =  \sum_{\alpha=1}^{+\infty  } I_{\alpha}^2 \vert I^L_{\alpha} \rangle \langle I^L_{\alpha} \vert
\label{EigenLaplacianDiff}
\end{eqnarray}
gives its spectral decomposition in terms of its positive eigenvalues $I_{\alpha}^2$
with the corresponding orthonormalized basis of eigenvectors $\vert I^L_{\alpha} \rangle $.
So the left singular kets $\vert I^L_{\alpha=0,1,..,+\infty} \rangle $
form an orthonormal basis of the space of scalar functions
 \begin{eqnarray}
\delta_{\alpha,\alpha'} && = \langle I^L_{\alpha} \vert I^L_{\alpha'} \rangle 
= \int_{\cal V}  d^3 \vec x \langle I^L_{\alpha} \vert \vec x \rangle \langle \vec x \vert I^L_{\alpha'} \rangle
\nonumber \\
\delta^{(3)}( \vec x - \vec x ')  && = \langle  \vec x \vert \vec x ' \rangle
 = \sum_{\alpha=0}^{ +\infty  } \langle  \vec x \vert I^L_{\alpha} \rangle \langle I^L_{\alpha} \vert \vec x ' \rangle
\label{orthoILdiff}
\end{eqnarray}

Whenever the basis of eigenvectors $\vert I^L_{\alpha=0,..,+\infty} \rangle $ 
of the opposite Laplacian $[- {\bold \Delta} ]$ of Eq. \ref{EigenLaplacianDiff}
over the bounded volume $\vec x \in {\cal V}$ is known, 
the corresponding right singular vectors $ \vert I^R_{\alpha=1,..,+\infty} \rangle$
can be obtained via the application of the operator ${\bold I}  =-  \vec{{\bold {grad}}}$
\begin{eqnarray}
{\bold I} \vert I^L_{\alpha} \rangle  &&  =-  \vec{{\bold {grad}}} \vert I^L_{\alpha} \rangle = I_{\alpha} \vert I^R_{\alpha} \rangle  \! \rangle
\label{RfromLdiff}
\end{eqnarray}
while the application of the adjoint operator ${\bold I}^{\dagger} = {\bold {div}}$ 
to these right singular vectors $ \vert I^R_{\alpha=1,..} \rangle$
\begin{eqnarray}
{\bold I}^{\dagger} \vert I^R_{\alpha} \rangle  \! \rangle  
&& =  {\bold {div}} \vert I^R_{\alpha=1,..} \rangle
= I_{\alpha} \vert I^L_{\alpha} \rangle 
\label{LfromRdiff}
\end{eqnarray}
reproduce the left singular vectors $\vert I^L_{\alpha} \rangle $.

The supersymmetric partner
\begin{eqnarray}
  {\bold I}  {\bold I}^{\dagger} = - {\bold {grad}} \ {\bold {div}}
 = - \begin{pmatrix} 
    \frac{\partial}{\partial x_1} 
 \\  
   \frac{\partial}{\partial x_2} 
 \\ 
   \frac{\partial}{\partial x_3} 
  \end{pmatrix}
  \begin{pmatrix} 
       \frac{\partial}{\partial x_1} 
 \  
   \frac{\partial}{\partial x_2} 
 \ 
   \frac{\partial}{\partial x_3} 
  \end{pmatrix}
  = - \bigg[  \frac{\partial^2}{\partial x_{\mu}\partial x_{\nu} }\bigg]_{\mu=1,2,3;\nu=1,2,3}
\label{partnerIncidencefokkerplanck}
\end{eqnarray}
corresponds to the opposite of the $3 \times 3$ symmetric matrix of the double derivatives $\frac{\partial^2}{\partial x_{\mu}\partial x_{\nu} } $.

The SVD decomposition of Eq. \ref{SVDI}
yields that its spectral decomposition 
\begin{eqnarray}
{\bold I} {\bold I}^{\dagger} && =  \sum_{\alpha=1}^{+\infty } I_{\alpha}^2 
\vert I^R_{\alpha} \rangle  \! \rangle \langle \! \langle I^R_{\alpha} \vert 
\label{SVDIdaggerIdiff}
\end{eqnarray}
involves the same strictly positive eigenvalues $I_{\alpha=1,..,+\infty}^2 >0$ 
as the opposite-Laplacian of Eq. \ref{EigenLaplacianDiff},
while the corresponding eigenvectors $ \vert I^R_{\alpha=1,..,} \rangle  \! \rangle $ are 
related to the eigenvectors $\vert I^L_{\alpha=1,..} \rangle $ of the opposite-Laplacian via Eq. \ref{RfromL}.
However the eigenvectors $ \vert I^R_{\alpha=1,..,} \rangle  \! \rangle $
should be supplemented by an infinite series of other kets
$\vert  I^R_{\alpha=0,-1,-2..,-\infty} \rangle \! \rangle $ 
(instead of the finite number $C$ of Eq. \ref{orthoIR} of the main text concerning Markov jump processes
in a space of $N$ configurations with $C$ independent cycles)
in order to obtain an orthonormal basis of the space of 3D vector fields 
  \begin{eqnarray}
\delta_{\alpha,\alpha'} && =  \langle \!\langle I^R_{\alpha} \vert I^R_{\alpha'} \rangle \! \rangle
\nonumber \\
{\bold 1}_{\{\text{3DVectorFields}\}} && = \sum_{\alpha=-\infty}^{+\infty }  \vert I^R_{\alpha} \rangle \! \rangle \langle \! \langle I^R_{\alpha} \vert
= \sum_{\alpha=1}^{+\infty  }  \vert I^R_{\alpha} \rangle \! \rangle \langle \! \langle I^R_{\alpha} \vert
+ \sum_{\alpha=-\infty}^{0  }  \vert I^R_{\alpha} \rangle \! \rangle \langle \! \langle I^R_{\alpha} \vert
\label{orthoIRdiff}
\end{eqnarray}
As discussed in detail in subsection \ref{subsec_discreteHelm} concerning the finite configuration space,
the right singular vectors $\vert I^R_{\alpha'} \rangle \! \rangle $
are directly related  to the Helmholtz decomposition for an arbitrary 3D vector fields $\vec v(\vec x)$
  \begin{eqnarray}
\vec v(\vec x) && = \sum_{\alpha=-\infty}^{+\infty} v_{\alpha}\vert I^R_{\alpha} \rangle  \! \rangle
\equiv \vec v ^{[I_.>0]}(\vec x)  + \vec v ^{[I_0=0]}(\vec x)
\nonumber \\
 \vec v ^{[I_.>0]}(\vec x) && \equiv  \sum_{\alpha=1}^{+\infty} v_{\alpha}\vert I^R_{\alpha} \rangle  \! \rangle  
\nonumber \\
\vec v ^{[I_0=0]}(\vec x) && \equiv \sum_{\alpha=-\infty}^{0} v_{\alpha}\vert I^R_{\alpha} \rangle  \! \rangle 
\label{Helmhotz}
\end{eqnarray}
into two orthogonal components with the following properties :

(i) the component $\vec v^{[I_0=0]}(\vec x) $ associated to the vanishing singular value $I_0=0 $
is annihilated by the divergence operator ${\bold I}^{\dagger}={\bold {div}} $
\begin{eqnarray}
0 =  {\bold I}^{\dagger}\vec v^{[I_0=0]}(\vec x)  =  {\bold {div}} \vec v^{[I_0=0]}(\vec x) 
\label{annidaggerI}
\end{eqnarray}
As already discussed on the special case of the divergencesless steady current $\vec j_*(\vec x)$
around Eq. \ref{divjnoneqfokkerplanck},
this divergencesless component $\vec v^{[I_0=0]}(\vec x) $
 can be rewritten as the curl of another divergenceless vector $\vec \omega ( \vec x ) $
\begin{eqnarray}
 \vec v^{[I_0=0]}( \vec x )  && = \vec \nabla \times \vec \omega ( \vec x )
 \nonumber \\
\bold{div} \  \vec \omega ( \vec x )  && =0
\label{vcurlomega}
\end{eqnarray}

(ii) the component  $ \vec v ^{[I_.>0]}(\vec x) $ associated to the strictly positive singular values $I_{\alpha}>0 $
can be rewritten using Eq. \ref{RfromLdiff}
  \begin{eqnarray}
 \vec v ^{[I_.>0]}(\vec x) && \equiv  \sum_{\alpha=1}^{+\infty} v_{\alpha}\vert I^R_{\alpha} \rangle  \! \rangle  
 = -  \vec{{\bold {grad}}} \left( \sum_{\alpha=1}^{+\infty} \frac{v_{\alpha}}{I_{\alpha} }
 \vert I^L_{\alpha} \rangle \right) \equiv -  \vec{{\bold {grad}}} \ g(\vec x)
\label{Helmhotzgrad}
\end{eqnarray}
as the opposite gradient of the scalar function
 \begin{eqnarray}
g(\vec x)  \equiv   \sum_{\alpha=1}^{+\infty} \frac{v_{\alpha}}{I_{\alpha} } \vert I^L_{\alpha} \rangle 
\label{Helmhotzgradg}
\end{eqnarray}

(iii) The application of the divergence  ${\bold I}^{\dagger}={\bold {div}} $ to the vector field $\vec v(\vec x) $ of Eq. \ref{Helmhotz} only involves the application to the gradient component $ \vec v ^{[I_.>0]}(\vec x) =-  \vec{{\bold {grad}}} \ g(\vec x)$ of Eq. \ref{Helmhotzgrad}
  \begin{eqnarray}
{\bold {div}} \vec v(\vec x) && = {\bold {div}} \vec v ^{[I_.>0]}(\vec x) +0 
= - {\bold \Delta } g(\vec x)
\label{Helmhotzdiv}
\end{eqnarray}
and reduces to the opposite Laplacian of the scalar function $g(\vec x) $.
The application of the ${\bold {curl}} $ to the vector field $\vec v(\vec x) $ of Eq. \ref{Helmhotz} only involves the application to the component $\vec v^{[I_0=0]}( \vec x )   = \vec \nabla \times \vec \omega ( \vec x )$
of Eq. \ref{vcurlomega}
  \begin{eqnarray}
\vec \nabla \times  \vec v(\vec x) && = 0  + \vec \nabla \times \vec v ^{[I_0=0]}(\vec x)
 \vec \nabla \times \left( \vec \nabla \times \vec \omega ( \vec x ) \right)
= \vec \nabla \left( \vec \nabla . \vec \omega ( \vec x ) \right) - \Delta \vec \omega ( \vec x )
= - \Delta \vec \omega ( \vec x )
\label{Helmhotzcurl}
\end{eqnarray}
and reduces to the opposite Laplacian of the the field $\vec \omega ( \vec x )$.

%%%%%%%%%%%%%%%%%%%%%%%%%%%%%%%%%%%%%%

\subsection{ Singular Value Decomposition of the current differential operator ${\bold J} $ 
and its adjoint $ {\bold J}^{\dagger}$ }

Let us replace the force $ \vec F(\vec x)=2 D(\vec x) \vec A(\vec x)$ in terms of the vector potential $\vec A(\vec x) $ introduced in Eq. \ref{Avectorpot} 
and write the Singular Value Decomposition for the current differential operator ${\bold J} $ 
and its adjoint $ {\bold J}^{\dagger}$ 
that involves an infinite series of strictly positive singular values $ \lambda_{\beta=1,2,..,+\infty} >0 $
 (instead of the $N$ values in Eq. \ref{SVDJ} of the main text concerning non-equilibrium Markov jump processes in a space of $N$ configurations)
\begin{eqnarray}
{\bold J} && =
  D(\vec x) \bigg( 2 \vec A(\vec x)   -  \vec \nabla \bigg) 
 =  \sum_{\beta=1}^{+\infty  } \lambda_{\beta} 
\vert \lambda^R_{\beta} \rangle  \! \rangle \langle \lambda^L_{\beta} \vert
\nonumber \\
{\bold J}^{\dagger} && \equiv    \bigg( 2 \vec A(\vec x)   + \vec \nabla \bigg) D(\vec x) . =  \sum_{\beta=1}^{+\infty  } \lambda_{\beta} 
\vert \lambda^L_{\beta} \rangle \langle \! \langle \lambda^R_{\beta} \vert
\label{SVDJdiff}
\end{eqnarray}

For a non-equilibrium steady state with a non-vanishing steady current $\vec j_*(\vec x) \ne 0$,
the left singular vectors $\lambda^L_{\beta=1,2,..,+\infty} $
form an orthonormal basis of the space of scalar functions on the finite domain ${\cal V} $
 \begin{eqnarray}
\delta_{\beta,\beta'} && =  \langle \lambda^L_{\beta} \vert \lambda^L_{\beta'} \rangle 
= \int_{\cal V} d^3 \vec x \langle \lambda^L_{\beta} \vert \vec x \rangle \langle \vec x \vert \lambda^L_{\beta'} \rangle
\nonumber \\
\delta^{(3)}( \vec x - \vec x ')  && = \langle  \vec x \vert \vec x ' \rangle
 = \sum_{\beta=1}^{ +\infty  } \langle  \vec x \vert \lambda^L_{\beta} \rangle \langle \lambda^L_{\beta} \vert \vec x ' \rangle
\label{orthoJLdiff}
\end{eqnarray}
that can be obtained from the spectral decomposition of the supersymmetric operator 
\begin{eqnarray}
 {\bold J}^{\dagger}{\bold J} && 
 = \bigg( 2 \vec A(\vec x)   + \vec \nabla \bigg) D^2(\vec x)  \bigg( 2 \vec A(\vec x)   -  \vec \nabla \bigg) 
=  \sum_{\beta=1}^{+\infty  } \lambda_{\beta}^2 \vert \lambda^L_{\beta} \rangle \langle \lambda^L_{\beta} \vert
\label{SVDdaggerJJdiff}
\end{eqnarray}

The supersymmetric partner
\begin{eqnarray}
  {\bold J}  {\bold J}^{\dagger}   =  \bigg[ D(\vec x) \bigg( 2  A_{\mu}(\vec x)   -   \frac{\partial}{\partial x_{\mu}} \bigg)  \bigg( 2  A_{\nu}(\vec x)   +  \frac{\partial}{\partial x_{\mu}} \bigg) D(\vec x) \bigg]_{\mu=1,2,3;\nu=1,2,3}
  = \sum_{\beta=1}^{+\infty  } \lambda^2_{\beta} 
\vert \lambda^R_{\beta} \rangle  \! \rangle \langle \langle \! \langle \lambda^R_{\beta} \vert
\label{partnerJJfokkerplanck}
\end{eqnarray}
involves the same strictly positive eigenvalues $\lambda^2_{\beta=1,2,...,+\infty}$ 
as Eq. \ref{SVDdaggerJJdiff},
where the corresponding eigenvectors $ \vert \lambda^R_{\beta} \rangle  \! \rangle $ are 
related to the eigenvectors $ \vert \lambda^L_{\beta} \rangle $ via
\begin{eqnarray}
{\bold J} \vert \lambda^L_{\beta} \rangle  &&  
= \lambda_{\beta} \vert \lambda^R_{\beta} \rangle  \! \rangle
\label{RfromLJdiff}
\end{eqnarray}
These eigenvectors $ \vert \lambda^R_{\beta=1,2,..,+\infty} \rangle  \! \rangle $
should be supplemented an infinite series of other kets
$\vert \lambda^R_{\beta=0,-1,..,-\infty} \rangle  \! \rangle $ 
(instead of the finite number $(C-1)$ of Eq. \ref{orthoJR} of the main text concerning Markov jump processes in a space of $N$ configurations with $C$ independent cycles)
in order to obtain an orthonormal basis of the space of 3D vector fields 
  \begin{eqnarray}
\delta_{\beta,\beta'} && =  \langle \!\langle \lambda^R_{\beta} \vert \lambda^R_{\beta'} \rangle \! \rangle
\nonumber \\
{\bold 1}_{\{\text{3DVectorFields}\}} && = \sum_{\beta=-\infty}^{+\infty } 
\vert \lambda^R_{\beta} \rangle \! \rangle \langle \! \langle \lambda^R_{\beta} \vert
=  \sum_{\beta=1}^{+\infty } 
\vert \lambda^R_{\beta} \rangle \! \rangle \langle \! \langle \lambda^R_{\beta} \vert
+  \sum_{\beta=-\infty}^{0 } 
\vert \lambda^R_{\beta} \rangle \! \rangle \langle \! \langle \lambda^R_{\beta} \vert
\label{orthoJRdiff}
\end{eqnarray}

As discussed in subsection \ref{subsec_helmForJ} for Markov jump processes, 
it is interesting to define the analog of the Helmholtz decomposition when the operator ${\bold I}= - {\bold {grad}}$ is replaced by the current operator ${\bold J}$ using its SVD decomposition of Eq. \ref{SVDJdiff} as follows.

An arbitrary 3D vector field $\vec u(\vec x)$ can be decomposed
into two orthogonal components
  \begin{eqnarray}
\vec u(\vec x) && = \sum_{\beta=-\infty}^{+\infty} u_{\beta}\vert \lambda^R_{\beta} \rangle  \! \rangle
\equiv \vec u ^{[\lambda_.>0]}(\vec x)  + \vec u^{[\lambda_0=0]}(\vec x)
\nonumber \\
 \vec u^{[\lambda_.>0]}(\vec x) && \equiv  \sum_{\beta=1}^{+\infty}   u_{\beta}\vert \lambda^R_{\beta} \rangle  \! \rangle
\nonumber \\
\vec u^{[\lambda_0=0]}(\vec x) && \equiv \sum_{\beta=-\infty}^{0} u_{\beta}\vert \lambda^R_{\beta} \rangle  \! \rangle
\label{HelmhotzJdiff}
\end{eqnarray}
 with the following properties :

(i) the component $\vec u^{[\lambda_0=0]}(\vec x) $ of Eq. \ref{HelmhotzJdiff}
associated to the vanishing singular value $\lambda_0=0 $
is annihilated by the adjoint operator ${\bold J}^{\dagger} $ of Eq. \ref{SVDJdiff}
\begin{eqnarray}
0 =  {\bold J}^{\dagger} \vec u^{[\lambda_0=0]}(\vec x) 
 =     \bigg( 2 \vec A(\vec x)   + \vec \nabla \bigg) .
 \bigg( D(\vec x)  \vec u^{[\lambda_0=0]}(\vec x) \bigg)
\label{annidaggerIdiff}
\end{eqnarray}

(ii) the component  $ \vec u ^{[\lambda_.>0]}(\vec x) $ 
associated to the strictly positive singular values $\lambda_{\beta=1,2,..+\infty}>0 $
can be rewritten using Eq. \ref{RfromLJdiff}
  \begin{eqnarray}
  \vec u^{[\lambda_.>0]}(\vec x) && =  \sum_{\beta=1}^{+\infty}   u_{\beta}\vert \lambda^R_{\beta} \rangle  \! \rangle
 =  {\bold J} \left( \sum_{\beta=1}^{+\infty} \frac{ u_{\beta} }{\lambda_{\beta}}  \vert \lambda^L_{\beta} \rangle \right)
 \equiv  {\bold J} \ k(\vec x)
\label{HelmhotzJdiffpos}
\end{eqnarray}
as the application of the current operator ${\bold J} $
to the scalar function
 \begin{eqnarray}
k(\vec x)  \equiv   \sum_{\beta=1}^{+\infty} \frac{ u_{\beta} }{\lambda_{\beta}}  \vert \lambda^L_{\beta} \rangle
\label{HelmhotzJdiffk}
\end{eqnarray}
  The application of  the adjoint operator ${\bold J}^{\dagger} $ to the vector field $\vec u(\vec x) $ of Eq. \ref{HelmhotzJdiff}
  only involves the application to the component $ \vec u^{[\lambda_.>0]}(\vec x) = {\bold J}  k(\vec x)$
of Eq. \ref{HelmhotzJdiffpos}
and reduces to the application of the supersymmetric operator $ {\bold J}^{\dagger} {\bold J}$ of Eq. \ref{SVDdaggerJJdiff}
to the scalar function $k(\vec x) $ of Eq. \ref{HelmhotzJdiffk}
  \begin{eqnarray}
{\bold J}^{\dagger} \vec u(\vec x) && = {\bold J}^{\dagger} \vec u^{[\lambda_.>0]}(\vec x) +0 
= {\bold J}^{\dagger} {\bold J}k(\vec x)
= \sum_{\beta=1}^{+\infty} 
 u_{\beta} \lambda_{\beta}  \vert \lambda^L_{\beta} \rangle 
\label{Jdaggu}
\end{eqnarray}

As discussed in subsection \ref{subsec_physical} concerning Markov jump processes, 
the component $\vert u^{[\lambda_0=0]} \rangle  \! \rangle $  
corresponds to the unphysical subspace orthogonal to the physical space
of the currents $\vec j_t (\vec x) =  {\bold J}   p_t (\vec x) $ 
that are obtained from 
the application of the current operator ${\bold J}$ to a ket $\vert p_t \rangle $.
The projector ${\bold {\cal P}}^{PhysicalSpaceCurrents} $ 
onto the physical subspace for currents
can be written either 
with the bi-orthogonal basis of the $\langle \! \langle  i_n  \vert $ and the $\vert j_n \rangle \! \rangle $ for $n=0,1,..,+\infty$
or with the orthonormalized basis $\vert \lambda^R_{\beta=1,..,+\infty} \rangle  \! \rangle $ of the right singular vectors of 
the current matrix ${\bold J}$ associated to the 
strictly positive singular values $\lambda^R_{\beta=1,..,+\infty} >0$
 \begin{eqnarray}
{\bold {\cal P}}^{PhysicalSpaceCurrents}   = \sum_{n=0}^{+\infty}    \vert j_n \rangle \! \rangle 
\langle \! \langle  i_n  \vert 
 = \sum_{\beta=1}^{+\infty  }  \vert \lambda^R_{\beta} \rangle \! \rangle \langle \! \langle \lambda^R_{\beta} \vert
\label{ProjectorPhysicalCurrentsDiff}
\end{eqnarray}
with the following consequences :

(a) As in Eqs \ref{jstarpstar} \ref{jstarprjSVDZero} \ref{jpstarjstar},
the decomposition of the steady current on the basis of right singular vectors $\vert \lambda^R_{\beta} \rangle  \! \rangle $ 
\begin{eqnarray}
\vert j_* \rangle \! \rangle = {\bold J} \vert p_* \rangle \equiv  \sum_{\beta=1}^{+\infty  } \lambda_{\beta} 
\vert \lambda^R_{\beta} \rangle  \! \rangle \langle \lambda^L_{\beta} \vert p_* \rangle
\label{Eqrhojsteady}
\end{eqnarray}
yields the vanishing of the components for $\beta=0,-1,..,- \infty $
\begin{eqnarray}
 \langle \! \langle \lambda^R_{\beta} \vert j_* \rangle \! \rangle  = 0
\ \ \ \text{ for } \ \ \beta=0,-1,..,- \infty
\label{jstarprjSVDZerodiff}
\end{eqnarray}
while the steady state $ \vert p_* \rangle$ can be obtained from the inversion of Eq. \ref{Eqrhojsteady}
\begin{eqnarray}
\vert p_* \rangle = {\bold J}^{pseudo[-1]} \vert j_* \rangle \! \rangle  
= \sum_{\beta=1}^{N  } \frac{1}{\lambda_{\beta} }
\vert \lambda^L_{\beta} \rangle \langle \! \langle \lambda^R_{\beta} \vert j_* \rangle \! \rangle
\label{jpstarjstardiff}
\end{eqnarray}
with the pseudo-inverse operator
\begin{eqnarray}
{\bold J}^{pseudo[-1]} \equiv  \sum_{\beta=1}^{+\infty  } \frac{1}{\lambda_{\beta} }
\vert \lambda^L_{\beta} \rangle \langle \! \langle \lambda^R_{\beta} \vert
\label{SVDJpseudodiff}
\end{eqnarray}
that can also be used for the excited right eigenvectors $ \vert r_{n=1,..,+\infty} \rangle $ 
 and their associated currents $ \vert j_n \rangle \! \rangle = {\bold J} \vert p_n \rangle$ as in Eq. \ref{rnpseudojn}
\begin{eqnarray}
\vert r_n \rangle = {\bold J}^{pseudo[-1]} \vert j_n \rangle \! \rangle  
= \sum_{\beta=1}^{+\infty  } \frac{1}{\lambda_{\beta} }
\vert \lambda^L_{\beta} \rangle \langle \! \langle \lambda^R_{\beta} \vert j_n \rangle \! \rangle
\label{rnpseudojndiff}
\end{eqnarray}

(b) As in Eqs \ref{spectrallsvdj} \ref{infromln}, the pseudo-inverse operator $ {\bold J}^{pseudo[-1]}$ of Eq. \ref{SVDJpseudodiff}
is also useful to invert 
\begin{eqnarray}
  \langle l_n \vert && =  \langle \! \langle i_n \vert {\bold J} 
 = 
  \sum_{\beta=1}^{+\infty  } \lambda_{\beta} 
\langle \! \langle i_n \vert \lambda^R_{\beta} \rangle  \! \rangle \langle \lambda^L_{\beta} \vert
\label{spectrallsvdjdiff}
\end{eqnarray}
into
 \begin{eqnarray}
\langle \! \langle  i_n  \vert 
&& = \sum_{\beta=1}^{+\infty  } \langle \! \langle  i_n  \vert  \vert \lambda^R_{\beta} \rangle \! \rangle 
\langle \! \langle \lambda^R_{\beta} \vert
= \sum_{\beta=1}^{+\infty  } \frac{ \langle  l_n  \vert   \lambda^L_{\beta} \rangle }{\lambda_{\beta} } 
\langle \! \langle \lambda^R_{\beta} \vert = \langle  l_n  \vert  {\bold J}^{pseudo[-1]}
\label{infromlndiff}
\end{eqnarray}
For $n=0$ with the left eigenvector $l_0(\vec x)=1$, Eq. \ref{infromlndiff}
yields that the bra $ \langle \! \langle  i_0  \vert$ is given by
 \begin{eqnarray}
\langle \! \langle  i_0  \vert && = \langle  l_0  \vert  {\bold J}^{pseudo[-1]}
= \sum_{\vec x}  \langle  l_0  \vert \vec x \rangle \langle \vec x \vert  {\bold J}^{pseudo[-1]}
\nonumber \\
&&  = \sum_{\vec x}    \langle \vec x \vert  {\bold J}^{pseudo[-1]}
= \sum_{\beta=1}^{+\infty  } \frac{ \langle  \vec x  \vert   \lambda^L_{\beta} \rangle }{\lambda_{\beta} } 
\langle \! \langle \lambda^R_{\beta} \vert 
\label{i0froml0diff}
\end{eqnarray}

 \subsection{ Discussion   }

In summary, the spectral decompositions of the non-hermitian Hamiltonians
${\bold H}={\bold I}^{\dagger}{\bold J}$  and ${\hat{\bold H}}= {\bold J}{\bold I}^{\dagger}$ 
as well as the properties of the Singular Value Decompositions of ${\bold I} $ and ${\bold J} $ 
that have been discussed in detail in the previous sections concerning non-equilibrium Markov jump processes in a space of $N$ configurations can be adapted 
to the case of non-equilibrium diffusion processes in dimension $d=3$ 
despite the technical differences.
The comparison between these two types of continuous-time Markov processes
is actually very useful to better understand each of them, since each particular idea 
is usually clearer or more familiar either in discrete space or in continuous space.

%%%%%%%%%%%%%%%%%%%%%%%%%%%%%%%%%%%%%%%%%%%%%%%%%%

%%%%%%%%%%%%%%%%%%%%%%%%%%%%%%%%%%%%%

\section{ Conclusions }

\label{sec_conclusion}

In order to clarify the spectral properties of continuous-time non-equilibrium Markov processes,
we have considered the continuity equation for the probability density as an Euclidean Schr\"odinger equation governed by the non-hermitian quantum Hamiltonian $ \bold{H}= {\bold {div}}  {\bold J} $ that is naturally factorized into the product of the divergence operator ${\bold {div}} $ and the current operator ${\bold J} $. 

We have first discussed in detail the case of non-equilibrium Markov jump processes in a space of $N$ configurations with $M$ links between them and $C=M-(N-1) \geq 1$ independent cycles. The factorization of the $N \times N$ non-hermitian Hamiltonian ${\bold H}= {\bold I}^{\dagger}  {\bold J} $ then involves the incidence matrix ${\bold I}$ and the current matrix ${\bold J} $ that are both of size $M \times N$, so that the supersymmetric partner ${\hat {\bold H} } = {\bold J}  {\bold I}^{\dagger}$ governing the dynamics of the currents living on the $M$ links is a priori of size $M \times M$. To better understand the relations between the spectral decompositions of these two Hamiltonians $ \bold{H}= {\bold I}^{\dagger}  {\bold J} $ and ${\hat {\bold H} } = {\bold J}  {\bold I}^{\dagger} $ with respect to their bi-orthogonal basis of right and left eigenvectors that characterize the relaxation dynamics towards the steady state and the steady currents, we have analyzed the properties of the Singular Value Decompositions of the two rectangular matrices ${\bold I}$ and ${\bold J} $ of size $M \times N $ and the interpretations in terms of discrete Helmholtz decompositions. This general framework
is illustrated by a simple translation-invariant example in Appendix \ref{app_example}.

Finally in the last section \ref{app_diffusion} of the main text, we have described how these spectral properties concerning non-equilibrium Markov jump processes in a space of $N$ configurations
can be translated for the case of non-equilibrium diffusion processes in dimension $d=3$,
where the two matrices ${\bold I}$ and ${\bold J} $ of size $M \times N $ become first-order differential operators acting on scalar functions to produce 3D vector fields, while the non-hermitian Hamiltonians ${\bold H}= {\bold I}^{\dagger}  {\bold J} $ and  ${\hat {\bold H} } = {\bold J}  {\bold I}^{\dagger}$ become second-order differential operators acting on scalar functions and 3D vectors respectively.

%%%%%%%%%%%%%%%%%%%%%%%%%%%%%%%%%%%%%%%%%%%

\appendix

\section{ Application to a translation-invariant ring with all-to-all transition rates depending on distance   }

\label{app_example}

In this Appendix, the general framework for arbitrary Markov jump processes described in the main text is illustrated by the example of a translation-invariant ring of $N$ sites where all-to-all transition rates depend only on the distance. As is well-known in spectral problems, the translation-invariance is a very strong symmetry that implies that eigenvectors reduce to Fourier modes that can be written explicitly for arbitrary sizes. This huge simplification will thus produce many additional specific properties with respect to the general theory discussed in the main text. 

%%%%%%%%%%%%%%%%%%%%%%%%%%%%%%%%%%%%

\subsection{ Model with $N=2L+1$ sites, $M= \frac{N(N-1)}{2}$ oriented links 
and $C= \frac{(N-1)(N-2)}{2}$ independent cycles }

The configuration space is a ring containing an odd number
\begin{eqnarray}
N=2L+1 \text{  of sites labelled by the positions} \ \ x=-L,-(L-1),..,-1,0,1,..,L-1,L
\label{xring}
\end{eqnarray}
with periodic boundary conditions $(L+1) \equiv -L$.
The model is fully-connected, i.e. transitions rates $w(.,.)$ exist between any pair of sites,
so that there are $M= \frac{N(N-1)}{2}=(2L+1) L$ oriented links
whose properties are described in the next subsection.

%%%%%%%%%%%%%%%%%%%%%%%%%%%%%%%%%%

\subsubsection{ Properties of the $M= \frac{N(N-1)}{2}=(2L+1) L$ oriented links }

The $M= \frac{N(N-1)}{2}=(2L+1) L= 2 L^2 + L$ oriented links between 
$x \in \{-L,+L \}$ and $x'=(x+y)$ with $y=1,..,L$ are characterized by the two transition rates of Eq. \ref{wrates}
where the two functions $D(x+y,x)=D_y$ and $A(x+y,x)=A_y$ depend only on the distance $y$
 \begin{eqnarray}
  w(x+y,x) && = D_y e^{A_y}
 \nonumber \\ 
 w(x,x+y) && = D_y e^{- A_y}
\label{wratesring}
\end{eqnarray}
so that the model is translation-invariant along the ring.

The current matrix ${\bold J}  $ of size $M \times N$ has for matrix elements
for $x \in \{-L,+L \}$ and $y=1,..,L$ with the rates of Eq. \ref{wratesring}
\begin{eqnarray}
\langle \! \langle _{\ x}^{x+y} \vert {\bold J} \vert z \rangle 
=  w(x+y,x) \delta_{z,x}  - w(x,x+y) \delta_{x+y,z} 
= D_y \left( e^{A_y} \delta_{z,x}  - e^{-A_y} \delta_{x+y,z} \right) 
\label{currentOpRing}
\end{eqnarray}
while the incidence matrix ${\bold I}$ of Eq. \ref{incidence}
reduces to
\begin{eqnarray}
\langle \! \langle _{\ x}^{x+y} \vert {\bold J} \vert z \rangle 
=   \delta_{z,x}  -  \delta_{x+y,z} 
\label{incidencering}
\end{eqnarray}

%%%%%%%%%%%%%%%%%%%%%%%%%%%%%%%%%%%%%%%%%%%%%%

\subsubsection{ Properties of the $C=M-(N-1) = \frac{(N-1)(N-2)}{2}=L (2L-1)$ independent cycles }

The $C=M-(N-1) = L (2L-1) = 2 L^2 -L$ independent cycles can be chosen as follows :

$\bullet$  the $(2L+1) (L-1)=2 L^2-L-1=C-1 $ cycles containing the oriented link $(x \to x+y)$ 
with the $(2L+1)$ values $x \in \{-L,+L \}$ and the $(L-1)$ values $y \in \{2,..,L\}$,
as well as the corresponding $y$ backward-links between nearest-neighbors along the ring
 \begin{eqnarray}
 \gamma(x+y,x) \equiv \{ x \to x+y \to x+y-1 \to  ... \to x+1 \to x\} \ \ \text{ of length} 
 \ \ l^{[\gamma(x+y,x)]}=(y+1) \in \{3,..,L+1\}
\label{gammaxrring}
\end{eqnarray}
are characterized by the following circulations of the antisymmetric function $A(.,.)$
\begin{eqnarray} 
\Gamma^{[\gamma(x+y,x)]}[A(.,.)]  = A_y - y A_1 \ \ \ \text{ for } \ \ x \in \{-L,+L \}
\ \ \text{and} \ \  y \in \{2,..,L\}
  \label{circulationAring}
\end{eqnarray}

$\bullet$ the remaining independent cycle can be chosen as 
the cycle containing the $N=(2L+1)$ nearest-neighbors links along the ring that will be denoted by
 \begin{eqnarray}
 \gamma_0 \equiv \{ 0 \to 1 \to ... \to +L \to -L \to ... \to -1 \to 0\} \ \ \text{ of length} 
 \ \ l^{[\gamma_0]}=N=2L+1
\label{gammaxrringzero}
\end{eqnarray}
with the circulation of the antisymmetric function $A(.,.)$
\begin{eqnarray} 
\Gamma^{[\gamma_0]}[A(.,.)]  = N  A_1 = (2L+1) A_1
  \label{circulationAzero}
\end{eqnarray}

As recalled around Eq. \ref{KolmogorovcirculationA}, the non-equilibrium nature of the steady state
comes from the non-vanishing of some of the circulations of Eqs \ref{circulationAring} and \ref{circulationAzero}.

%%%%%%%%%%%%%%%%%%%%%%%%%%%%%%%%%%%

\subsection{ Steady currents $ j_*(x+y,x) $ on the $M$ oriented links and the decomposition onto the $C$ independent cycles }

Since the $(2L+1)$ sites are equivalent as a consequence of the translation-invariance of the model,
the normalized steady state is simply uniform
\begin{eqnarray} 
p_*(x)  = \frac{1}{2L+1}  
  \label{steadyring}
\end{eqnarray}
The corresponding steady currents on the $M$ oriented links 
can be computed from the rates of Eq. \ref{wratesring} 
 for $x \in \{-L,+L \}$ and $x'=(x+y)$ with $y=1,..,L$
\begin{eqnarray}
 j_*(x+y,x) = - j_*(x,x+y) && =   w(x+y,x) p_*(x) - w(x,x+y) p_*( x+y ) 
  \nonumber \\
 && = \frac{ D_y }{2L+1} \left[ e^{A_y}  - e^{-A_y}  \right] =  \frac{ D_y 2 \sinh(A_y) }{2L+1} 
\label{currentsteadyring}
\end{eqnarray}
that depend only on $y$ as a consequence of translation-invariance, and that are non-vanishing
when $A_y \ne 0$.

On the other hand, these steady currents on the $M$ oriented links
 can be rewritten via Eq. \ref{jsteadycyclesGamma}
in terms of the $C$ steady cycle-currents $j_*^{Cycle[\gamma] }$ flowing around the $C$ independent cycles $\gamma$ of Eqs \ref{gammaxrring} \ref{gammaxrringzero} for $x \in \{-L,+L \} $
\begin{eqnarray} 
 j_* (x+y,x) &&  =  j_*^{Cycle[\gamma(x+y,x)]} \ \ \ \text{ for } \ \  \ \  y \in \{2,..,L\}
 \nonumber \\
 j_*(x+1,x) && = j_*^{Cycle[\gamma_0]} -  \sum_{y=2}^L \sum_{z=x+1-y}^x j_*^{Cycle[\gamma(z+y,z)]}
  \label{jsteadycyclesGammaring}
\end{eqnarray}
The identification with Eq. \ref{currentsteadyring} yields that the cycle-currents $j_*^{Cycle[\gamma(x+y,x)]} $ for $ y \in \{2,..,L\}$
reduce to the link-currents $j_* (x+y,x) $
\begin{eqnarray} 
j_*^{Cycle[\gamma(x+y,x)]}= j_* (x+y,x) &&  = \frac{ D_y 2 \sinh(A_y) }{2L+1}   \ \ \ \text{ for } \ \  \ \  y \in \{2,..,L\}
  \label{jsteadycyclesGammaringiden}
\end{eqnarray}
while the remaining cycle current along the ring reads
\begin{eqnarray} 
 j_*^{Cycle[\gamma_0]} =  j_*(x+1,x)  +  \sum_{y=2}^L \sum_{z=x+1-y}^x j_*^{Cycle[\gamma(z+y,z)]}
 = \frac{ D_1 2 \sinh(A_1) }{2L+1} +  \sum_{y=2}^L y \frac{ D_y 2 \sinh(A_y) }{2L+1}
  \label{jsteadycyclesGammaringidenzero}
\end{eqnarray}

%%%%%%%%%%%%%%%%%%%%%%%%%%%%%%%%%%%%%%%%

\subsection{ Spectral properties governing the dynamics of the probabilities $p_t(x) $ and of the currents $j_t(x',x)$ }

%%%%%%%%%%%%%%%%%%%%%%%%%%%%%%%%%%%%%%%%

\subsubsection{ Reminder on the diagonalization of $N \times N$ real-space-circulant-matrices in the Fourier basis}

A real-space-circulant matrix ${\bold C} $ 
whose matrix elements only depend only on the difference $(x'-x)$
\begin{eqnarray}
\langle x' \vert {\bold C} \vert x \rangle = c(x'-x)
\label{Circulantmatrix}
\end{eqnarray}
becomes diagonal
\begin{eqnarray}
{\bold C}  =  \sum_{q=-L}^{+L  } {\hat c}_q \vert q \rangle \langle q \vert
\label{circulantfourier}
\end{eqnarray}
in the orthonormalized Fourier basis $\vert q \rangle$ 
involving the $N=2L+1$ values $q=-L,..,+L$ with components
\begin{eqnarray} 
\langle x \vert q \rangle  \equiv  \frac{1}{\sqrt{2L+1}} e^{i 2 \pi \frac{q x}{2L+1}  }
  \label{fourierbasis}
\end{eqnarray}
with the eigenvalues
\begin{eqnarray} 
 {\hat c}_q = \sum_{y=-L}^L c(y) e^{-i 2 \pi \frac{q y}{2L+1}  }
  \label{circulanteigen}
\end{eqnarray}
In the three following subsections, this general result will be applied 
 to the three following $N \times N$ real-space-circulant matrices 
 that appear in the present translation-invariant model :
the non-hermitian Hamiltonian $\bold{H}={\bold I}^{\dagger}{\bold J} $, 
and the two supersymmetric matrices ${\bold J}^{\dagger}{\bold J} $ 
and ${\bold I}^{\dagger}{\bold I} $.

%%%%%%%%%%%%%%%%%%%%%%%%%%%%%%%%%%%%%%%%

\subsubsection{ Spectral decomposition of the opposite Markov generator 
$(-{\bold w})={\bold H}$ in the Fourier basis }

With the rates of Eq. \ref{wratesring},
the off-diagonal elements of Eq. \ref{Hwrates} read for $y \in \{1,..,L\} $
 \begin{eqnarray}
 {\bold H} (x+y,x) &&  = - D_y e^{ A_y}   \equiv h(y) 
   \nonumber \\ 
 {\bold H} (x,x+y) && =- D_y e^{-A_y}  \equiv h(-y) 
\label{Hwratesring}
\end{eqnarray} 
while the diagonal elements of Eq. \ref{Hwdiag}
become
\begin{eqnarray}
{\bold H} (x,x)=    \sum_{y=1}^L   \left[ D_y e^{- A_y}   + D_y e^{ A_y}  \right] \equiv h(0)
\label{Hwdiagring}
\end{eqnarray}

This non-hermitian real-space-circulant matrix ${\bold H} $ is thus diagonal in the 
orthonormalized Fourier basis $\vert q \rangle $ of Eq. \ref{fourierbasis},
which is an important simplification with respect to the bi-orthogonal basis of left and right eigenvectors 
for the general case of Eq. \ref{spectral}
\begin{eqnarray}
{\bold H} && =  \sum_{q=-L}^{+L  } E_q \vert q \rangle \langle q \vert
\label{Hcirculantfourier}
\end{eqnarray}

The eigenvalues $ E_q $ are given by Eq. \ref{circulanteigen} in terms of the function $h(y)$ 
introduced in Eqs \ref{Hwratesring} \ref{Hwdiagring}
\begin{eqnarray} 
E_q && = \sum_{y=-L}^L h(y) e^{-i 2 \pi \frac{q y}{2L+1}  }
= h(0) + \sum_{y=1}^L \left[ h(y) e^{-i 2 \pi \frac{q y}{2L+1}  }+h(-y) e^{i 2 \pi \frac{q y}{2L+1}  }\right]
\nonumber \\
&& = \sum_{y=1}^L D_y \left[  e^{ A_y} \left( 1- e^{-i 2 \pi \frac{q y}{2L+1}  }  \right)
 +  e^{- A_y} \left(1- e^{i 2 \pi \frac{q y}{2L+1}  }   \right) \right] 
  \label{circulanteigenH}
\end{eqnarray}
Besides the expected vanishing eigenvalue $E_{(q=0)}=0$ associated to zero-momentum $q=0$,
 the other $(2L)$ eigenvalues $E_{q\ne 0} $ for $q=-L,..,-1,+1,..,=L$ 
display the following real parts and imaginary parts
(the real parts being strictly positive in agreement with the general property of Eq. \ref{RealEn}) 
 \begin{eqnarray}
&& {\text {Re}}(E_{q }) =\sum_{y=1}^L D_y
 \left[  e^{ A_y} +  e^{- A_y}  \right] \left[ 1- \cos \left( 2 \pi \frac{q y}{2L+1} \right)  \right]
  = {\text {Re}}(E_{-q })>0 \ \ \ \text{ for } \ \ q=1,2,..,L
\nonumber \\
&& {\text {Im}}(E_q) =\sum_{y=1}^L D_y \left[  e^{ A_y} -  e^{- A_y}  \right]
\sin \left( 2 \pi \frac{q y}{2L+1} \right)  = - {\text {Im}}(E_{-q })\ \ \ \text{ for } \ \ q=1,2,..,L
\label{RealImEq}
\end{eqnarray}
and thus appear in complex-conjugate pairs $(E_q;E_{-q})$ for $q=1,2,..,L$
 \begin{eqnarray}
E_{-q}=  {\text {Re}}(E_{-q })+i  {\text {Im}}(E_{-q })
=  {\text {Re}}(E_{q })-i  {\text {Im}}(E_q) = E_q^*
\label{ComplexConjugateEq}
\end{eqnarray}

These $(2L)$ non-vanishing eigenvalues $E_{q\ne 0} $ govern the relaxation 
towards the normalized uniform steady state $ p_*(x)  = \frac{1}{2L+1}  $ of Eq. \ref{steadyring}
via the propagator of Eq. \ref{propagator}
\begin{eqnarray} 
p_t(x \vert x_0) \equiv \langle x \vert e^{-t {\bold H}  } \vert x_0 \rangle 
&& =  \sum_{q=-L}^L  e^{ - t E_q } \langle x \vert q \rangle \langle q \vert x_0 \rangle
= \frac{1}{2L+1} \sum_{q=-L}^L  e^{ - t E_q }  e^{i 2 \pi \frac{q (x-x_0) }{2L+1}  } 
 \nonumber \\
&&  
= \frac{1}{2L+1} \left[ 1 + \sum_{q=1}^L  \left( e^{ - t E_q }  e^{i 2 \pi \frac{q (x-x_0) }{2L+1}  } 
+ e^{ - t E_{-q} }  e^{-i 2 \pi \frac{q (x-x_0) }{2L+1}  }  \right) \right]
 \label{propagatorring}
\end{eqnarray}
that is real, as it should, as a consequence of the complex-conjugate property $E_{-q}= E_q^* $ of Eq. \ref{ComplexConjugateEq}.

%%%%%%%%%%%%%%%%%%%%%%%%%%%%%%%%%%%%%%%%%%%

\subsubsection{ Singular Value Decomposition of the Current matrix ${\bold J} $ of size $M \times N$ }

The $N \times N$ supersymmetric matrix $ {\bold J}^{\dagger}{\bold J} $ 
with the matrix elements of Eq. \ref{jdaggerj}
\begin{eqnarray}
 \langle x \vert {\bold J}^{\dagger} {\bold J}  \vert x' \rangle  
= \begin{cases}
\displaystyle \sum_{y=1}^L \left( w^2(x+y,x)+w^2(x-y,x) \right)
=\sum_{y=1}^L D_y^2 \left( e^{2 A_y} + e^{- 2 A_y} \right)  \ \ \ \ \ \ \ \ \ \ \  \text{   if }  x=x'  
 \\
 - D^2_{\vert x'-x \vert} \ \ \ \ \ \ \ \ \ \ \ \ \text{   if $x \ne x'$ }  
\end{cases}
\label{jdaggerjring}
\end{eqnarray}
is a symmetric real-space-circulant-matrix that becomes diagonal 
in the orthonormalized Fourier basis $\vert q \rangle $ of Eq. \ref{fourierbasis}
\begin{eqnarray}
 {\bold J}^{\dagger}{\bold J} && 
=  \sum_{q=-L}^{+L  } \lambda_{q}^2 \vert q \rangle \langle q \vert
\label{SVDdaggerJJring}
\end{eqnarray}
where the $(2L+1)$ eigenvalues given by Eq. \ref{circulanteigen} in terms of the matrix elements of Eq. \ref{jdaggerjring}
\begin{eqnarray}
 \lambda_q^2 && =\sum_{y=1}^L D_y^2 \left( e^{2 A_y} + e^{- 2 A_y} \right)
-  \sum_{y=1}^L D_y \left(  e^{-i 2 \pi \frac{q y}{2L+1}  } +e^{i 2 \pi \frac{q y}{2L+1}  } \right)
\nonumber \\
&&  = \sum_{y=1}^L D_y^2 \left[  e^{2 A_y} + e^{- 2 A_y} - 2 \cos \left( 2 \pi \frac{q y }{2L +1} \right) \right]
\label{SVDdaggerJJeigen}
\end{eqnarray}
are strictly positive for any non-equilibrium case where the $A_{y=1,2,..,L}$ do not all vanish.

So the left singular vectors $\vert \lambda^L_q \rangle $ that appear in the spectral decomposition
of $ {\bold J}^{\dagger}{\bold J} $ in Eq. \ref{SVDdaggerJJ}
reduce to the Fourier basis of Eq. \ref{fourierbasis}
\begin{eqnarray} 
\vert \lambda^L_q \rangle= \vert q \rangle  
  \label{fourierbasisSVDJ}
\end{eqnarray}
and the Singular Value Decomposition of the current matrix ${\bold J} $ of Eq. \ref{SVDJ}
reads
\begin{eqnarray}
{\bold J} && =  \sum_{q=-L}^{+L  } \lambda_q 
\vert \lambda^R_{q} \rangle  \! \rangle \langle q \vert
\nonumber \\
{\bold J}^{\dagger} && =  \sum_{q=-L}^{+L  } \lambda_q 
\vert q \rangle   \langle \! \langle \lambda^R_{q} \vert
\label{SVDJring}
\end{eqnarray}
where the corresponding right singular vectors $\vert \lambda^R_{q} \rangle  \! \rangle $ can be obtained via Eq. \ref{RfromLJ} using Eq. \ref{currentOpRing}
\begin{eqnarray}
\langle \! \langle _{\ x}^{x+y} \vert \lambda^R_q \rangle  \! \rangle 
&& 
= \frac{1}{  \lambda_q } \langle \! \langle _{\ x}^{x+y} \vert  {\bold J} \vert q \rangle    
=  \frac{1}{  \lambda_q } \sum_{z=-L}^{+L} \langle \! \langle _{\ x}^{x+y} {\bold J} \vert z \rangle  
\langle z \vert q \rangle  
% \nonumber \\ &&
=  \frac{1}{  \lambda_q \sqrt{2L+1}} \sum_{z=-L}^{+L}
 D_y \left( e^{A_y} \delta_{z,x}  - e^{-A_y} \delta_{x+y,z} \right) e^{i 2 \pi \frac{q z}{2L+1}  }
 \nonumber \\
&&=  \frac{1}{  \lambda_q \sqrt{2L+1}} 
 D_y e^{i 2 \pi \frac{q x}{2L+1}  }
 \left( e^{A_y}    - e^{-A_y}  e^{i 2 \pi \frac{q y }{2L+1}  } \right) 
\label{RfromLJring}
\end{eqnarray}
and satisfy the orthonormalization of Eq. \ref{orthoJR} 
 \begin{eqnarray}
  \langle \!\langle \lambda^R_{q} \vert \lambda^R_{q'} \rangle \! \rangle
&&=  \sum_{x=-L}^{L} \sum_{y=1}^L    \langle \!\langle \lambda^R_{q} \vert _{x}^{x+y} \rangle \! \rangle
\langle \! \langle _{x}^{x+y} \vert \lambda^R_{q'} \rangle \! \rangle
\nonumber \\
&& =  \frac{1}{  \lambda_q \lambda_{q'}} 
\left[ \frac{1}{ (2L+1)} \sum_{x=-L}^{L}   e^{i 2 \pi \frac{(q' -q) x}{2L+1}  } \right]
\sum_{y=1}^L    D_y^2 
 \left( e^{A_y}    - e^{-A_y}  e^{-i 2 \pi \frac{q y }{2L+1}  } \right) 
 \left( e^{A_y}    - e^{-A_y}  e^{i 2 \pi \frac{q' y }{2L+1}  } \right)
\nonumber \\
&& =  \frac{\delta_{q,q'} }{  \lambda_q^2 } 
 \sum_{y=1}^L D_y^2 \left[  e^{2 A_y} + e^{- 2 A_y} - e^{i 2 \pi \frac{q y }{2L+1}  } - e^{-i 2 \pi \frac{q y }{2L+1}  } \right]
 = \delta_{q,q'}
\label{orthoJRring}
\end{eqnarray}
where the last simplification comes from the values $\lambda_q^2 $ of Eq. \ref{SVDdaggerJJeigen}.
 
%%%%%%%%%%%%%%%%%%%%%%%%%%%%%%%%%%%%%%%%%%%

\subsubsection { Singular Value Decomposition of the Incidence matrix ${\bold I} $ of size $M \times N$ }

Since the incidence matrix ${\bold I} $ can be obtained from the current matrix ${\bold J} $ when 
$D_y \to 1$ and $A_y \to 0$, one can adapt the results of the previous subsection concerning ${\bold J} $ as follows.

The supersymmetric matrix $ {\bold I}^{\dagger}{\bold I} $ corresponding to the opposite Laplacian
with the matrix elements of Eq. \ref{incidenceidagger}
\begin{eqnarray}
 \langle x \vert {\bold I}^{\dagger} {\bold I}  \vert x' \rangle  
= \begin{cases}
2L  \ \ \ \ \ \ \ \ \ \ \  \text{   if }  x=x'  
 \\
 - 1 \ \ \ \ \ \ \ \ \ \ \  \text{   if $x \ne x'$ }  
\end{cases}
\label{idaggeriring}
\end{eqnarray}
leads to the spectral decomposition in the Fourier basis
\begin{eqnarray}
 {\bold I}^{\dagger}{\bold I} && 
=  \sum_{q=-L}^{+L  } I_q^2 \vert q \rangle \langle q \vert
\label{{SVDdaggerII}ring}
\end{eqnarray}
where the eigenvalues corresponds to Eq. \ref{SVDdaggerJJeigen}
when 
$D_y \to 1$ and $A_y \to 0$
\begin{eqnarray}
 I_q^2  = \sum_{y=1}^L  2 \left[  1 -  \cos \left( 2 \pi \frac{q y }{2L +1} \right) \right]
\label{SVDdaggerIIeigen}
\end{eqnarray}
with the vanishing eigenvalue $I_{q=0}=0$ associated to zero-momentum $q=0$ in agreement with
the general property of Eq. \ref{Izero}.

So the left singular vectors $\vert I^L_q \rangle $ of the incidence matrix
reduce to the Fourier basis of Eq. \ref{fourierbasis}
\begin{eqnarray} 
\vert I^L_q \rangle= \vert q \rangle  
  \label{fourierbasisSVDI}
\end{eqnarray}
and the Singular Value Decomposition of the incidence matrix ${\bold I} $ of Eq. \ref{SVDI}
that only involves the $(N-1)=2L$ positive singular values $ I_q>0 $ associated to $q \ne 0$
reads
\begin{eqnarray}
{\bold I} && =  \sum_{q \in \{-L,..,+L\} \text{and} \ q \ne 0}  I_q \vert I^R_{q} \rangle  \! \rangle \langle q \vert
\nonumber \\
{\bold I}^{\dagger} && =  \sum_{q \in \{-L,..,+L\} \text{and} \ q \ne 0}  I_q 
\vert q \rangle   \langle \! \langle I^R_{q} \vert
\label{SVDIring}
\end{eqnarray}
where the corresponding right singular vectors $\vert I^R_{q \ne 0} \rangle $ can be obtained via Eq. \ref{RfromL} using Eq. \ref{incidencering}
\begin{eqnarray}
\langle \! \langle _{\ x}^{x+y} \vert I^R_q \rangle  \! \rangle 
&& 
= \frac{1}{  I_q } \langle \! \langle _{\ x}^{x+y} \vert {\bold I} \vert q \rangle  
=  \frac{1}{  I_q } \sum_{z=-L}^{+L} \langle \! \langle _{\ x}^{x+y} {\bold I} \vert z \rangle  
\langle z \vert q \rangle    
=   \frac{1}{  I_q \sqrt{2L+1}} 
  e^{i 2 \pi \frac{q x}{2L+1}  }
 \left( 1    -   e^{i 2 \pi \frac{q y }{2L+1}  } \right) 
\label{RfromLiring}
\end{eqnarray}

The orthonormalization of Eq. \ref{orthoIR} can be checked as in Eq. \ref{orthoJRring}
when $D_y \to 1$ and $A_y \to 0$
  \begin{eqnarray}
  \langle \!\langle I^R_{q} \vert I^R_{q'} \rangle \! \rangle
= \delta_{q,q'}
\label{orthoIRing}
\end{eqnarray}

It is also useful to compute the scalar products with the 
right singular vectors $\vert \lambda^R_{q} \rangle  \! \rangle $ of Eq. \ref{RfromLJring}
 \begin{eqnarray}
  \langle \!\langle I^R_{q} \vert \lambda^R_{q'} \rangle \! \rangle
&&=  \sum_{x=-L}^{L} \sum_{y=1}^L    \langle \!\langle I^R_{q} \vert _{x}^{x+y} \rangle \! \rangle
\langle \! \langle _{x}^{x+y} \vert \lambda^R_{q'} \rangle \! \rangle
\nonumber \\
&& =  
 \frac{1}{  I_q  \lambda_{q'}  } 
\left[ \frac{1}{ (2L+1)} \sum_{x=-L}^{L}   e^{i 2 \pi \frac{(q' -q) x}{2L+1}  } \right]
 \sum_{y=1}^L    D_y 
 \left( 1    -   e^{-i 2 \pi \frac{q y }{2L+1}  } \right) 
 \left( e^{A_y}    - e^{-A_y}  e^{i 2 \pi \frac{q' y }{2L+1}  } \right) 
\nonumber \\
&& =   \frac{\delta_{q,q'}}{  I_q  \lambda_{q}  } 
 \sum_{y=1}^L D_y \left[  e^{ A_y} \left( 1- e^{-i 2 \pi \frac{q y}{2L+1}  }  \right)
 +  e^{- A_y} \left(1- e^{i 2 \pi \frac{q y}{2L+1}  }   \right) \right] 
 = \frac{E_q}{  I_q  \lambda_{q}  }  \delta_{q,q'}
\label{orthoIJRring}
\end{eqnarray}
where the last simplification comes from the eigenvalues $E_q $ of Eq. \ref{circulanteigenH}.
These scalar products appear when one plugs
the SVD of ${\bold I}^{\dagger} $ from Eq. \ref{SVDIring} 
and the SVD of ${\bold J} $ 
from Eq. \ref{SVDJring}
into the Hamiltonian ${\bold H} $
\begin{eqnarray}
{\bold H}={\bold I}^{\dagger}{\bold J} && =
\left(  \sum_{q \in \{-L,..,+L\} \text{and} \ q \ne 0}  I_q 
\vert q \rangle   \langle \! \langle I^R_{q} \vert \right)
\left(  \sum_{q'=-L}^{+L  } \lambda_{q'} 
\vert \lambda^R_{q'} \rangle  \! \rangle \langle q' \vert \right)
\nonumber \\
&& = \sum_{q \in \{-L,..,+L\} \text{and} \ q \ne 0}  E_q \vert q \rangle   \langle q \vert 
\label{HSVDIJring}
\end{eqnarray}
and one recovers the spectral decomposition of Eq. \ref{Hcirculantfourier}
as it should for consistency.

%%%%%%%%%%%%%%%%%%%%%%%%%%%%

\subsubsection{ Spectral decomposition of the supersymmetric partner ${\hat{\bold H}}= {\bold J}{\bold I}^{\dagger}$ governing the dynamics of the currents}

As explained in detail the main text, the time-dependent currents $  j_t(x+y,x)$ defined on the $M=N-1+C$ oriented links live in the smaller subspace of dimension $N=2L+1$ associated to the projector of Eq.\ref{ProjectorPhysicalCurrents} that reads for the present model
 \begin{eqnarray}
{\bold {\cal P}}^{PhysicalSpaceCurrents}   = \sum_{q=-L}^{+L}    \vert j_q \rangle \! \rangle 
\langle \! \langle  i_q  \vert 
 = \sum_{q=-L}^{L  }  \vert \lambda^R_q \rangle \! \rangle \langle \! \langle \lambda^R_q \vert
\label{ProjectorPhysicalCurrentsRing}
\end{eqnarray}
where the second expression involve the right singular vectors $\vert \lambda^R_{q} \rangle $ appearing in the SVD of the current matrix ${\bold J}$ that have been written in Eq. \ref{RfromLJring}, while the first expression is related to the spectral decomposition 
of the supersymmetric partner ${\hat{\bold H}}^{Phys}$ governing the dynamics of the physical currents in Eq. \ref{jtpropaini}
\begin{eqnarray}
e^{-t {\hat{\bold H}}^{Phys}}= \sum_{q=-L}^L  e^{ - t E_q }
\vert j_q \rangle \! \rangle \langle \! \langle  i_q  \vert 
\label{partnerHring}
\end{eqnarray}
The general properties of $\vert j_q \rangle \! \rangle $ and $ \langle \! \langle  i_q  \vert $
have been discussed in detail between Eqs \ref{jtpropa} and \ref{jtpropaini}.
In the present translation-invariant model governed by the Fourier modes $\vert q \rangle$, the ket $\vert j_q \rangle \! \rangle \equiv {\bold J} \vert q \rangle$  
given by Eq. \ref{jn} is directly related to the right singular ket $\vert \lambda^R_q \rangle  \! \rangle $ of Eq. \ref{RfromLJring}
\begin{eqnarray}
\vert j_q \rangle \! \rangle && \equiv {\bold J} \vert q \rangle = \lambda_q  \vert \lambda^R_q \rangle  \! \rangle
\label{spectraljring}
\end{eqnarray}
with the components 
\begin{eqnarray}
\langle \! \langle _{\ x}^{x+y}\vert j_q \rangle \! \rangle && = 
 \frac{1}{   \sqrt{2L+1}} 
 D_y e^{i 2 \pi \frac{q x}{2L+1}  }
 \left( e^{A_y}  
  - e^{-A_y}  e^{i 2 \pi \frac{q y }{2L+1}  }
 \right) 
\label{spectraljringcompo}
\end{eqnarray}
while $\langle \! \langle  i_q  \vert$ 
is given by Eq. \ref{infromln} that involves the pseudo-inverse $  {\bold J}^{pseudo[-1]}$
\begin{eqnarray}
\langle \! \langle  i_q  \vert && =  \langle  q  \vert  {\bold J}^{pseudo[-1]}
=  \langle  q  \vert \left(  \sum_{q'=-L}^{+L  }  
\frac{ \vert q' \rangle   \langle \! \langle \lambda^R_{q'} \vert }{ \lambda_{q'} } \right) 
= \frac{ \langle \! \langle \lambda^R_{q} \vert }{ \lambda_{q} }
\label{infromlnring}
\end{eqnarray}
and is thus also directly related to the right singular vector $\langle \! \langle \lambda^R_{q} \vert $.
Eqs \ref{spectraljring} and \ref{infromlnring} yields the simple relation
 \begin{eqnarray}
   \vert j_q \rangle \! \rangle \langle \! \langle  i_q  \vert 
 =   \vert \lambda^R_q \rangle \! \rangle \langle \! \langle \lambda^R_q \vert
\label{ProjectorPhysicalCurrentsRingfouriercompare}
\end{eqnarray}
so that the two decompositions of the projector of Eq. \ref{ProjectorPhysicalCurrentsRing}
coincide term by term in the present translation-invariant model governed by Fourier modes.
As a consequence, the spectral decomposition of Eq. \ref{partnerHring}
for the non-hermitian Hamiltonian $ {\hat{\bold H}}^{Phys}$ can be also rewritten in terms of
the orthonormalized basis of the right singular vectors $\vert \lambda^R_{q} \rangle $ of the current matrix ${\bold J}$
\begin{eqnarray}
e^{-t {\hat{\bold H}}^{Phys}}= \sum_{q=-L}^L  e^{ - t E_q }
 \vert \lambda^R_q \rangle \! \rangle \langle \! \langle \lambda^R_q \vert
\label{partnerHringortho}
\end{eqnarray}

%%%%%%%%%%%%%%%%%%%

 \subsection{ Discussion   }

 In conclusion, the translation-invariance of the model yields that the Fourier basis $\vert q \rangle$ of the configuration space plays an essential role in the spectral properties of
 the various matrices and produces many simplifications with respect to the general theory described in the main text.

%%%%%%%%%%%%%%%%%%%%%%%%%%%%%%%%%%%%%%%%%%%%%%%%%
%
%%%%%%%%%%%%%%%%%%%%%%%%%%%%%%%%%%%%%%%%%%

\end{document}